\documentclass[aps,pre,reprint,groupedaddress,longbibliography,nofootinbib]{revtex4-2}

\usepackage{amsmath,amssymb,mathtools,bbm,bm}
\usepackage{graphicx}
\usepackage{booktabs}
\usepackage{microtype}
\usepackage[colorlinks=true,linkcolor=blue,citecolor=blue,urlcolor=blue]{hyperref}
\usepackage{float}
\usepackage{adjustbox}
\usepackage{subcaption}

\newcommand{\E}{\mathbb{E}}

\newcommand{\dd}{\mathrm{d}}
\newcommand{\1}{\mathbf{1}}

\newcommand{\Part}{\mathcal{P}}

\allowdisplaybreaks

\begin{document}

\title{Maximum Entropy Temporal Networks}

\author{Paolo Barucca}
\address{Department of Computer Science, University College London}

\begin{abstract}
Temporal networks consist of timestamped directed interactions that may appear continuously in time, yet few studies have directly tackled the continuous-time modeling of networks. Here, we introduce a maximum-entropy approach to temporal networks and with basic assumptions on constraints, the corresponding network ensembles admit a modular and interpretable representation: a set of global time processes and a static maximum-entropy edge, e.g. node pair, probability. This time-edge labels factorization yields closed-form log-likelihoods, degree, clustering and motif expectations, and yields a whole class of effective generative models. 
We provide the maximum-entropy derivation for the non-homogeneous Poisson Process (NHPP) intensities governing the probability of directed edges in temporal networks via the functional optimization over path entropy, connecting NHPP modeling to maximum-entropy network ensembles. 
NHPPs consistently improve log-likelihood over generic Poisson processes, while the maximum-entropy edge labels recover strength constraints and reproduce expected unique-degree curves.
We discuss the limitations of this framework and how it can be integrated with multivariate Hawkes calibration procedures, renewal theory, and neural kernel estimation in graph neural networks. 
\end{abstract}

\maketitle

\section{Introduction}
Networks are widely used to describe the architecture of social, biological, and economic systems. While static network models have been studied extensively, many real-world systems are better represented by temporal networks, where interactions are time-stamped and evolve continuously over time. In such systems, links appear as discrete events on a fixed or evolving node set, producing temporal structures that cannot be captured by static snapshots alone. A central challenge is to develop models for temporal networks that are both interpretable and predictive, while resting on principled statistical foundations.

In the static setting, maximum-entropy ensembles have become a cornerstone of network science. Inspired by Jaynes’ work on information-theoretic inference~\cite{Jaynes1957a}, maximum-entropy network ensembles reproduce prescribed node- or network-level statistics while remaining otherwise maximally random. These ensembles have been successfully applied to reproduce degree and strength distributions, higher-order constraints, and mesoscale structures, thereby providing a baseline against which empirical patterns can be tested~\cite{cimini2019statistical,newman2018networks}. 
Similarly, the identification of communities and mesoscopic organization through block models has led to principled clustering frameworks that combine interpretability with prediction accuracy~\cite{karrer2011stochastic,Rohe2016CoclusteringDG,newman2001structure}. Together, these works have established maximum-entropy principles and related exponential-family ensembles as indispensable for static network modeling \cite{cimini2019statistical}.

At the same time, temporal dynamics display signatures that cannot be explained by static constraints alone. 
A wide body of research has shown that interaction streams exhibit burstiness, memory, and correlation patterns across scales, leading to non-Poissonian inter-event time distributions~\cite{Masuda2016,holme2012temporal,vazquez2007impact}.
Such deviations from Poissonian timing can often be modeled through renewal processes \cite{hiraoka2020modeling}, where inter-event times are independent but follow heavy-tailed distributions such as power-law, lognormal, or Weibull forms~\cite{goh2008burstiness,karsai2012correlated,cox1962renewal}.
Renewal processes naturally reproduce the alternation of rapid bursts and long inactive periods observed in empirical data, providing a minimal, interpretable baseline for bursty temporal networks that captures temporal heterogeneity without invoking explicit event–event dependencies.
In contrast, self-exciting point processes, introduced in Hawkes’ seminal work~\cite{hawkes1971point}, describe causal or feedback-driven event generation, reproducing clustered activity and contagion-like excitation effects in contexts ranging from seismology to financial order flows~\cite{Ogata1988,Bacry2015,bacry2016first,filimonov2012quantifying, seabrook2026modelling}.

Temporal networks naturally combine both aspects: structural constraints in the marks (who interacts with whom) and temporal patterns in the timing of events. Empirical studies have shown that such systems exhibit distinctive motifs, heterogeneous attention, and limited memory~\cite{Masuda2016,clemente2023temporal}. Recent work has introduced maximum-entropy approaches for discrete-time temporal networks~\cite{clemente2023temporal}, while advances in fitting multivariate Hawkes processes have produced highly calibrated generative models~\cite{bacry2016first,Achab2017,Soliman2022}. Yet, despite these parallel developments, there is no principled maximum-entropy ensemble for temporal networks in continuous time. In particular, there is no framework that jointly encodes structural constraints and realistic bursty dynamics, while remaining analytically tractable and interpretable.

Here we introduce such a framework. By maximizing path entropy for non-homogeneous Poisson processes under structural and temporal constraints, we derive a family of maximum-entropy ensembles that factorize into a \emph{time} component—modeled by non-homogeneous Poisson intensities—and a \emph{mark} component—given by static maximum-entropy edge weights subject to strengths, masks, or block constraints. 
This separation yields closed-form log-likelihoods, analytical expectations for degrees and unique edges, and motif statistics that combine mark probabilities with temporal cross-dependencies. Empirically, we show on a benchmark dataset (Enron email dataset) that adding a time layer consistently improves likelihoods -yet not necessarily achieving optimality in terms of information criteria- and motif fits over Poisson baselines, while maximum-entropy marks reproduce degree and clustering. 
Our contribution unifies maximum-entropy ensembles with bursty temporal processes, filling a key gap in the modeling of temporal networks and offering a tractable foundation for future extensions involving calibrated kernels, motif analysis, and neural inference methods.

\section{Methods}
Any temporal network observed over a time window $[0,T]$ can be 
represented as a marked point process. Each interaction is an event 
occurring at a time $t_k \in [0,T]$ and carrying a mark 
$m_k = (i \to j)$ that identifies the ordered pair of nodes involved. 
Thus the dataset consists of events 
$\{(t_k,m_k)\}_{k=1}^K$, where $K$ is the total number of timestamped edges.
We model the event times as a non–homogeneous Poisson process (NHPP). 
A NHPP on $[0,T]$ with intensity $\lambda(t)\ge 0$ is a counting process $N(t)$ such that:
\begin{itemize}
    \item  for any measurable set $A\subset[0,T]$, the number of events 
$N(A)$ is Poisson distributed with mean 
$\Lambda(A)=\int_A \lambda(t)\,dt$;
\item for disjoint sets $A_1,\dots,A_k$, the random variables 
$N(A_1),\dots,N(A_k)$ are independent.
\end{itemize}
Equivalently, $\lambda(t)$ is the instantaneous event rate, so that 
$\lambda(t)dt$ gives the probability of observing an event in 
$[t,t+dt)$ to first order in $dt$.
In the marked setting, we introduce edge–specific intensities 
$\lambda_{ij}(t)\ge 0$, describing the instantaneous rate at which the 
directed edge $(i,j)$ occurs at time $t$. 
The global event rate is therefore
\[
\Lambda(t) = \sum_{(i,j)\in E} \lambda_{ij}(t),
\]
where $E$ denotes the set of admissible directed edges on the node set $V$.
Under this model, the log–likelihood of observing the events 
$\{(t_k,m_k)\}_{k=1}^n$, with $m_k=(i_k\to j_k)$, is
\begin{equation}
\ell(\{\lambda_{ij}\})
=
\sum_{k=1}^K \log \lambda_{i_k j_k}(t_k)
-
\sum_{(i,j)\in E}\int_0^T \lambda_{ij}(t)\,dt.
\end{equation}
The integrated intensity $\mu_{ij}=\int_0^T \lambda_{ij}(t)\,dt=N_{ij}$ corresponds to the expected number of interactions from node $i$ to node $j$, i.e., the expected directed edge weight in the time-aggregated network. 
Node strengths are recovered as $s_i^{out}=\sum_j \mu_{ij}$ and 
$s_j^{in}=\sum_i \mu_{ij}$, linking the temporal model directly to standard weighted network quantities.
Taking expectation with respect to the process yields the Poisson entropy functional
\[
\mathcal{H}[\lambda]
=
-\int_0^T
\big(
\lambda(t)\log\lambda(t)-\lambda(t)
\big)\,dt.
\]

Maximizing entropy thus corresponds to selecting intensity functions that are as unstructured as possible subject to chosen constraints on the expected values over a set of prescribed functions of the events, in direct analogy to the canonical ensemble in statistical mechanics.

The central methodological step is to impose both temporal and structural constraints on the intensities. Temporal constraints specify aggregate time profiles for sets of edges, e.g. the total rate of activity from a given sender, within a block pair, or globally. Structural constraints fix time–integrated quantities such as edge totals, node strengths, or block sums. Formally, if $\Part=\{E_r\}$ is a partition of the edge set and $G_r(t)$ are prescribed profiles, then we require $\sum_{(i,j)\in E_r}\lambda_{ij}(t)=G_r(t)$ for all $r$ and $t$. Similarly, integrated constraints take the form $\int_0^T \lambda_{ij}(t)\,dt=\mu_{ij}=N_{ij}$, for edge constraints, $\sum_j\mu_{ij}=s_i^{\rm out}$, for node-strengths constraints, or $\sum_{i\in a,j\in b}\mu_{ij}=N_{ab}$, for block-pair flows. 
A block partition is defined as a mapping $g:V\to B$ assigning each node to one of $|B|$ groups. 
Block-pair quantities $N_{ab}$ aggregate expected interactions between nodes in block $a$ and block $b$.
These linear restrictions capture both prior information and desired marginal statistics of the temporal network.

To enforce these conditions, we introduce a Lagrangian functional combining the entropy with time-dependent multipliers $\alpha_r(t)$ for the temporal partitions and static multipliers for the mark-side margins. Explicitly,
\begin{align}
\mathcal{L}=\;&-\sum_{(i,j)\in E}\int_0^T\!\big(\lambda_{ij}\log\lambda_{ij}-\lambda_{ij}\big)\,dt \nonumber\\
&+\sum_r\int_0^T \alpha_r(t)\!\left(\sum_{(i,j)\in E_r}\lambda_{ij}(t)-G_r(t)\right)\,dt \nonumber\\
&+\sum_{(i,j)}\Theta_{ij}\!\left(\int_0^T\lambda_{ij}(t)\,dt-N_{ij}\right),
\end{align}
and similar terms for node strengths and block constraints. Stationarity with respect to each $\lambda_{ij}(t)$ yields
\begin{equation}
\log \lambda_{ij}(t)=\alpha_{r(i,j)}(t)+\Psi_{ij}-1,
\end{equation}
with $\Psi_{ij}$ collecting all static multipliers (edge, sender, receiver, block) and mask terms. 
Thus, the solution always factorizes as
\begin{equation}
\lambda_{ij}(t)=\phi_{r(i,j)}(t)\,w_{ij}, \qquad \phi_r(t)=e^{\alpha_r(t)},\;\;w_{ij}=e^{\Psi_{ij}-1}.
\end{equation}
This is the key structural result: time and marks separate cleanly. Each temporal partition has its own profile $\phi_r(t)$, while all static constraints determine the weights $w_{ij}$. The separation is reminiscent of exponential-family factorization and ensures convexity of the inference problem.

The factorization allows direct enforcement of constraints. For each partition, $\phi_r(t)$ is pinned by $G_r(t)$ up to a normalization, while the integrated quantities $\mu_{ij}=\Phi_{r(i,j)} w_{ij}$ with $\Phi_r=\int_0^T\phi_r(t)\,dt$ must match the prescribed margins. The latter reduces to masked biproportional scaling (iterative proportional fitting) on $w_{ij}$, a well-studied convex optimization that generalizes the construction of static maximum-entropy ensembles. Importantly, this procedure remains feasible under masks that restrict admissible edges or enforce block quotas, provided the margin sums are consistent.

Several notable cases emerge. If only edge totals are fixed and no temporal profile is prescribed, one obtains the edge–Poisson model with constant rates $\lambda_{ij}(t)=\mu_{ij}^\star/T$. If instead a global Hawkes intensity $\lambda_H(t)$ is imposed as an non-homogeneous Poisson process, the model factorizes as $\lambda_{ij}(t)=(\mu_{ij}^\star/\Lambda_H)\,\lambda_H(t)$, effectively rescaling the Hawkes process across edges. In Figure \ref{fig:enron-lambda-GH} we show the empirical and fitted global Hawkes intensity for the Enron dataset.
Adding node strengths with a mask yields a temporal analogue of the Directed Weighted Configuration Model, while imposing block-degree quotas enforces a fixed number of unique edges per block pair. 
Finally, partitioning by sender and assigning sender-specific profiles recovers a flexible class of non-homogeneous but still convex ensembles. 
In each case, sampling reduces to simulating the partition-level time processes and assigning marks according to $w_{ij}$, thereby generalizing classical shuffling methods to bursty, constraint-aware temporal null models.

\begin{figure}
\includegraphics[width=0.5\textwidth]{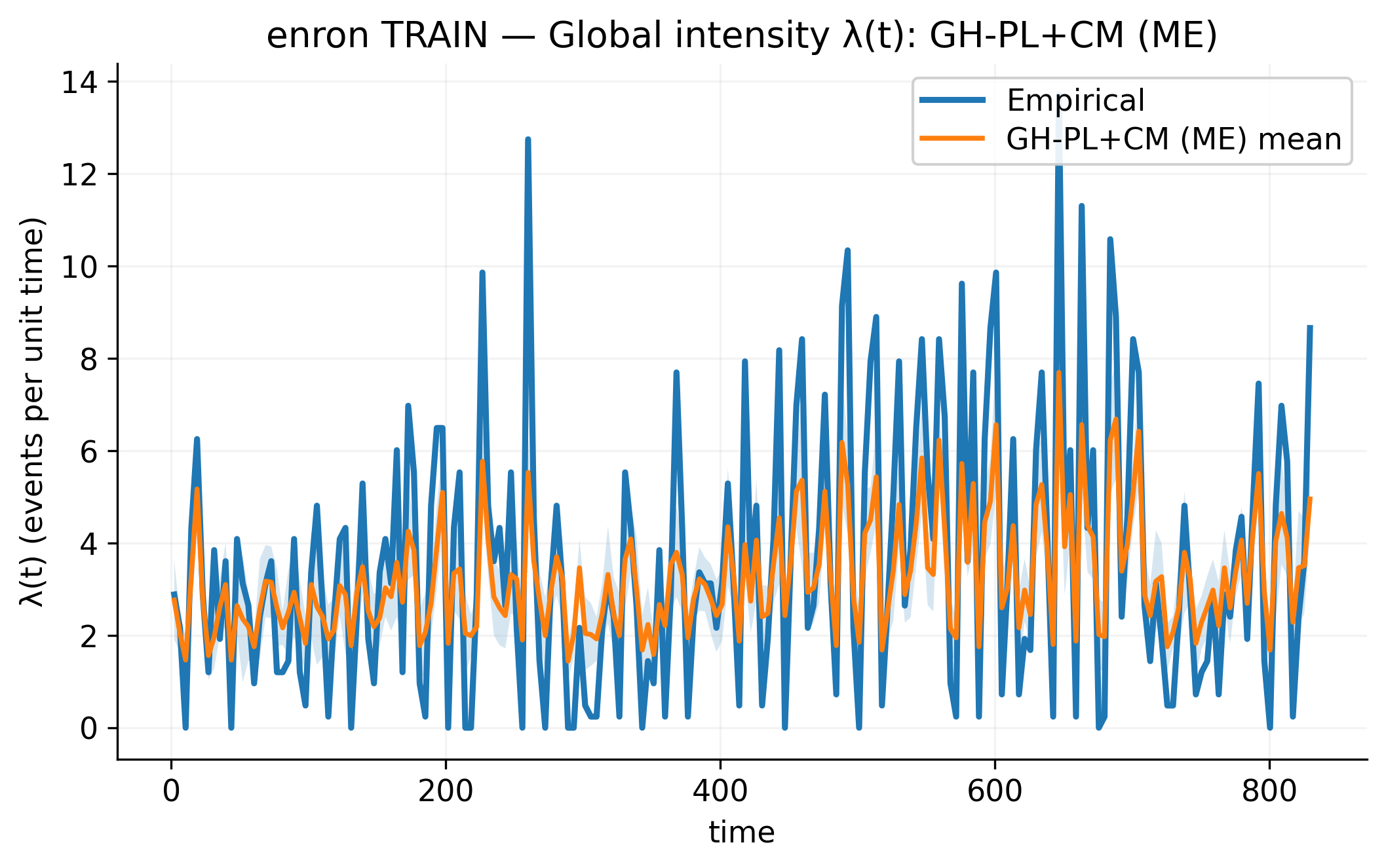}
\caption{Enron. Intensity over time obtained by fitting one global Hawkes Process to the whole sequence of timestamps.}
\label{fig:enron-lambda-GH}
\end{figure}

\subsection{Network properties}
The separation $\lambda_{ij}(t)=\phi_{r(i,j)}(t)\,w_{ij}$ permits a wide range of statistics to be computed in closed form, providing a tractable description of structural, temporal, and higher-order properties of temporal networks. 
At the structural level, the integrated intensity $\mu_{ij}=\Phi_{r(i,j)}w_{ij}$ defines the expected number of events along edge $(i,j)$ in $[0,T]$, with $\Phi_r=\int_0^T\phi_r(t)\,dt$, see \ref{fig:enron-matrices}.

Summing over targets or sources yields the expected out-strength $s_i^{\rm out}=\sum_j \mu_{ij}$ and in-strength $s_j^{\rm in}=\sum_i\mu_{ij}$, while block sums $\mu_{ab}=\sum_{i\in a,j\in b}\mu_{ij}$ provide mesoscopic statistics. 
The expected number of distinct edges activated is
\begin{equation}
\E[U_{\to}]=\sum_{i\neq j}\big(1-e^{-\mu_{ij}}\big),
\end{equation}
i.e. one minus the probability of the edge never being active in the time-window considered, generalizing the coverage formulas of static ensembles and allowing direct comparison with observed unique-edge counts. 
These statistics show how structural constraints can be consistently enforced on average in the ensemble, similarly to the corresponding static ensembles.

Turning to the temporal dimension, the partition profiles $\phi_r(t)$ govern inter-event times and burstiness. For a homogeneous Poisson process with rate $\bar\lambda$, inter-event times are exponentially distributed with variance $1/\bar\lambda^2$. In contrast, when $\phi_r(t)$ follows a Hawkes process with branching ratio $\eta<1$ and baseline intensity $\mu_r$, the stationary rate is $\bar\lambda_r=\mu_r/(1-\eta)$, and the variance is inflated by the kernel-dependent autocovariance $\gamma_r(u)=\E(\lambda_r(t),\lambda_r(t+u))$. 
This produces over-dispersion in inter-event intervals, a hallmark of bursty temporal dynamics. 
Within the ensemble, each partition inherits this property, so burstiness is naturally captured by the time-layer while marks remain maximally random subject to constraints. 
Thus, our framework reproduces both static marginals and temporal dispersion statistics in a unified manner.

Most importantly, the simple combination of mark probabilities and temporal covariances enables explicit computation of motif statistics. Consider ordered event pairs within a lag window $(0,\Delta]$ and an observation interval $I=[a,b]$. For each partition $r$, define the mass of the rows and columns of the edge weights $r^{(r)}_i=\sum_j \Pi^{(r)}_{ij}$ and $c^{(r)}_j=\sum_i \Pi^{(r)}_{ij}$, where $\Pi^{(r)}_{ij}=w_{ij}/W_r$ is the normalized edge distribution, where $W_r:=\sum_{(i,j)\in E_r} w_{ij}$.

\begin{figure}[t]
  \includegraphics[width=0.5\textwidth]{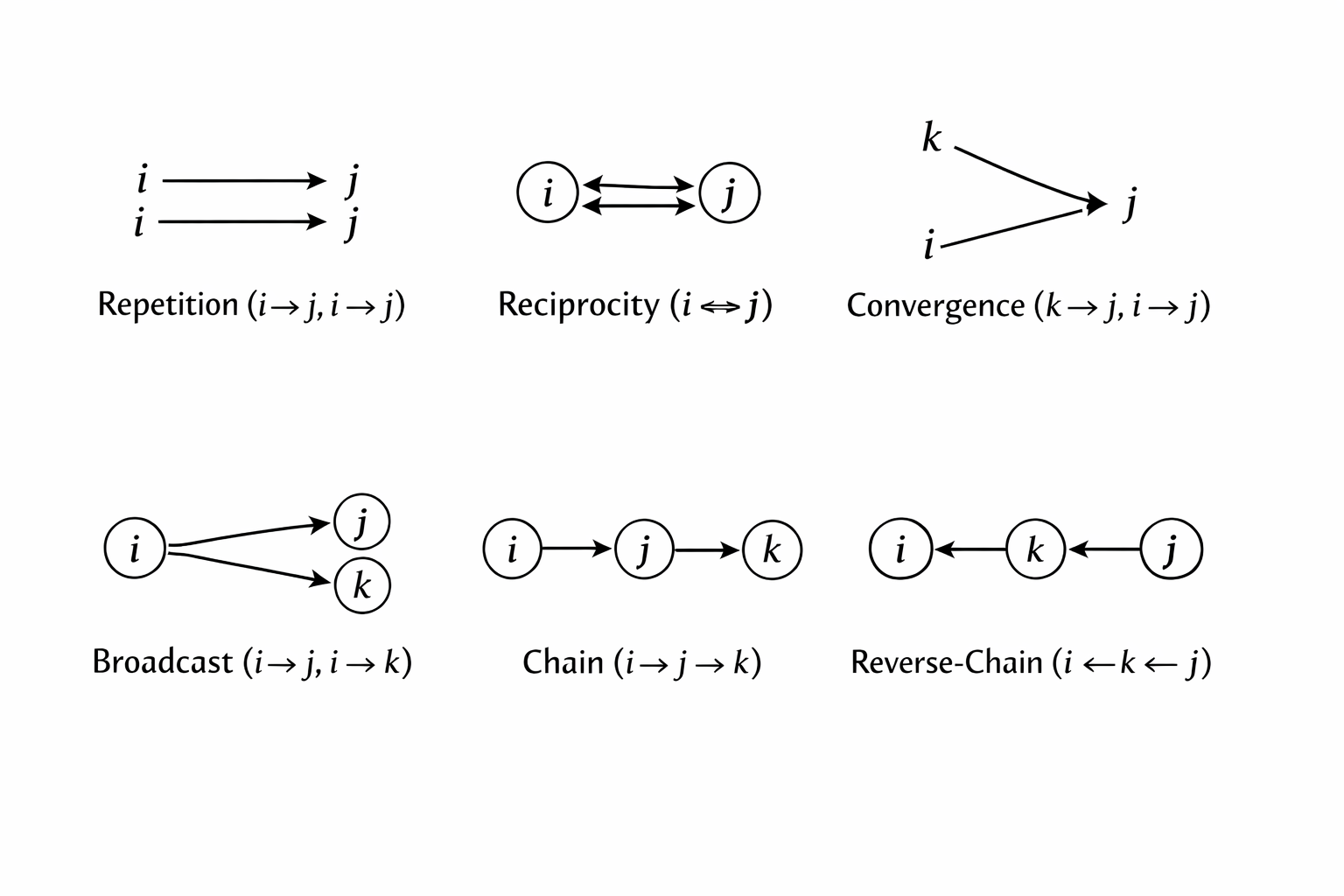}
  \caption{Directed two-event temporal motifs involving at most three nodes. 
  From left to right: \emph{repeat} ($i\!\to\!j$, $i\!\to\!j$), 
  \emph{reciprocity} ($i\!\to\!j$, $j\!\to\!i$), 
  \emph{convergence} ($i\!\to\!j$, $k\!\to\!j$), 
  \emph{broadcast} ($i\!\to\!j$, $i\!\to\!k$), 
  \emph{chain} ($i\!\to\!j$, $j\!\to\!k$), and 
  \emph{reverse-chain} ($i\!\to\!j$, $k\!\to\!i$). 
  These six configurations form a mutually exclusive and exhaustive set of 
  two-event motifs including at most three distinct nodes.
  }
  \label{fig:motifs}
\end{figure}

Then we can evaluate the pure-edge probabilities of two-edge motifs, in particular,  we are showing the repeat case, and reciprocity case, and the broadcast and convergence case:

\begin{align}
p_{\mathrm{REP}}^{(r,s)} &= \sum_{i,j} \Pi^{(r)}_{ij} \, \Pi^{(s)}_{ij}, \\[4pt]
p_{\mathrm{REC}}^{(r,s)} &= \sum_{i,j} \Pi^{(r)}_{ij} \, \Pi^{(s)}_{ji}, \\[4pt]
p_{\mathrm{BCAST}}^{(r,s)} &= \sum_i r^{(r)}_i \, r^{(s)}_i \;-\; p_{\mathrm{REP}}^{(r,s)}, \\[4pt]
p_{\mathrm{CONV}}^{(r,s)} &= \sum_j c^{(r)}_j \, c^{(s)}_j \;-\; p_{\mathrm{REP}}^{(r,s)},
\end{align}

where repeat considers the repetition of the same edge, reciprocity the appearance of both the connection from $i$ to $j$ and from $j$ to $i$, the broadcast case the appearance of two (distinct) edges with the same source $i$, and finally in the convergence case, node $j$ receives two (distinct) edges.
This is the full list of 2-edge motifs involving 2 and 3 nodes in which time-ordering is irrelevant: if we accounted for time-ordering we could consider the chain and reverse-chain (see Fig.\ref{fig:motifs}).

These quantities are purely structural, i.e. edge-dependent yet time-independent, reflecting the allocation of probability mass across edges.
Temporal amplification is captured by the cross-moment
\begin{equation}
C_{r,s}(\Delta;I)=\int_a^b \int_0^\Delta \E[\lambda_r(t)\lambda_s(t+u)]\,du\,dt,
\end{equation}
where $\lambda_r(t)=\sum_{(i,j)\in E_r}\lambda_{ij}(t)$. For $r\neq s$ independent, $C_{r,s}=|I|\Delta \bar\lambda_r\bar\lambda_s$, while for $r=s$ one obtains $C_{r,r}=|I|(\bar\lambda_r^2\Delta+\int_0^\Delta \gamma_r(u)\,du)$. Combining both pieces yields the general motif master formula:
\begin{equation}
\E\big[M^{(\Delta)}_{\mathrm{TYPE}}(I)\big]=\sum_{r,s} p^{(r,s)}_{\mathrm{TYPE}}\,C_{r,s}(\Delta;I).
\end{equation}

This decomposition explicitly shows that temporal autocorrelation $\gamma_r$ amplifies short-lag motifs, while mark probabilities control the balance among motif types. In particular, self-exciting Hawkes kernels enhance broadcast and convergence motifs in line with empirical bursty communication, while static constraints can enforce degree and clustering profiles.

Together, these results demonstrate the analytical tractability of the ensemble. Edge-level expectations follow from $\mu_{ij}$, burstiness is quantified by inter-event variance and autocovariance, and motif statistics emerge from a combination of structural and temporal factors.

\section{Results} 
A central contribution of our framework is to provide a principled benchmark against which empirical temporal network properties can be evaluated for statistical significance. By specifying both structural constraints (degrees, strengths, block totals) and temporal constraints (global or partitioned profiles, Hawkes kernels, renewal processes), the maximum-entropy ensembles define a family of null models that preserve low-order features while randomizing higher-order patterns. 

\begin{figure*}[t]
  \centering
  \includegraphics[width=1\textwidth]{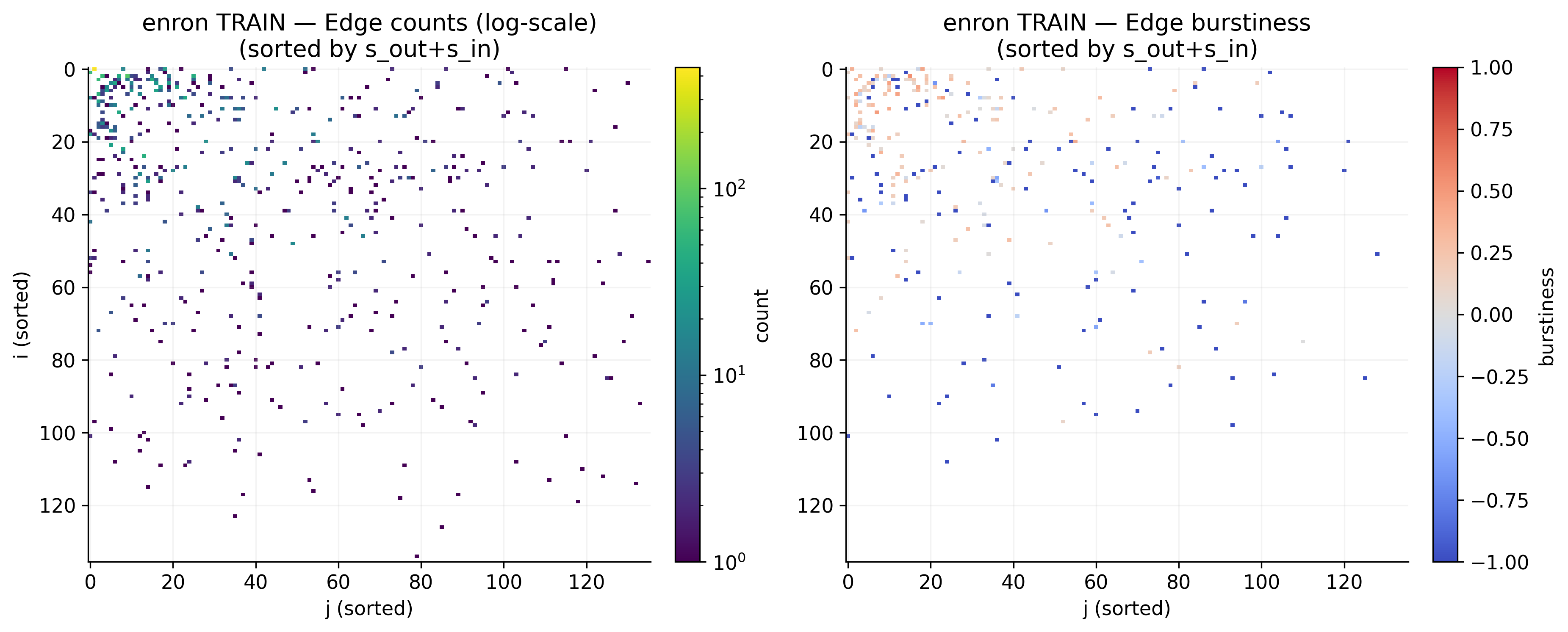}
  \caption{ Log of event counts per directed edge (left) and burstiness index per edge (right) for the Enron dataset. Nodes are sorted by total strength $s_i^{\mathrm{out}} + s_i^{\mathrm{in}}$, and edges are shown in the resulting order. The train split denotes the temporal segment of the data used for parameter calibration.}
  \label{fig:enron-matrices}
\end{figure*}

This allows one to compute the expected values of edge statistics, burstiness, and motifs under the null model and to compare them directly with observed values. Any significant deviation between empirical and expected quantities can thus be interpreted as evidence for structure beyond the imposed constraints, analogously to the role of canonical ensembles in equilibrium statistical mechanics.

We tested this principle across four widely studied temporal network datasets and report results on the Enron mail corpus. 
The Enron corpus records timestamped internal communications among employees of the Enron Corporation over a period of years. This dataset has become a benchmark for studies of temporal communication patterns because it exhibits both strong structural heterogeneity (hubs of activity, communities of practice) and highly bursty temporal dynamics. 
Figure \ref{fig:enron-matrices} provides an analysis of the temporal and degree heterogeneity of the Enron dataset, where edges are sorted by total strength $s_i^{\text{out}} + s_i^{\text{in}}$. The left panel shows the logarithm of event counts per directed edge, revealing strong structural heterogeneity: a small set of high-activity nodes concentrate a large fraction of interactions, while most dyads remain sparsely populated. The right panel reports the burstiness index for each edge, highlighting substantial temporal variability even among edges with comparable total counts. Highly active edges are not uniformly bursty; rather, burstiness varies across dyads, indicating that structural intensity and temporal irregularity are  partially decoupled.

The challenge is to separate the features of the dataset —such as degree heterogeneity or daily cycles— from significant temporal signatures, such as anomalous motifs, excess reciprocity, or collective bursts, by defining ensembles that enforce the known marginals (e.g., strengths, block structure) and temporal baselines (Poisson or Hawkes profiles) while leaving higher-order correlations unconstrained. By generating expectations and variances for the relevant statistics under these ensembles, we can define the associated empirical deviations.

We test that the frequency of reciprocated motifs largely exceeds the maximum-entropy expectation given node strengths and bursty Hawkes timing. 
This can be simply interpreted as people working at Enron replying to each other more than would be expected simply from being active and emailing in bursts, thus suggesting genuine conversational feedback between workers. 
Figure \ref{fig:enron-motifs-large} reports the time evolution of the four directed motifs under $\Delta_{\text{large}}$ for the Enron TRAIN split, comparing empirical ratios with the maximum-entropy temporal ensembles. Reciprocity (REC) and repetition (REP) are substantially underestimated by configuration-only models, even when coupled with Hawkes timing, indicating that strength constraints and global burstiness alone are insufficient to reproduce dyadic memory effects. Introducing block structure partially corrects this gap at short times but still decays away from the empirical curve, suggesting that mesoscale structure explains part—but not all—of the excess reciprocity. For convergence (CON) and broadcast (BRO), Hawkes-based models improve the early-time amplification relative to Poisson baselines, yet persistent discrepancies remain, especially at longer horizons where empirical ratios stabilise above the ensemble means. Overall, the figure highlights that while the time–mark factorization captures bursty temporal activity and degree constraints, motif persistence in the data reflects additional temporal correlations beyond those encoded by global Hawkes excitation and static block masks.
The ensemble serves as the baseline against which significance is judged: if an observed property falls within the distribution generated by the maximum-entropy temporal model, it is explained by the enforced constraints; if not, it signals mechanisms beyond those constraints.

\begin{figure*}[t]
  \centering
  \begin{subfigure}{0.48\textwidth}
    \includegraphics[width=\linewidth]{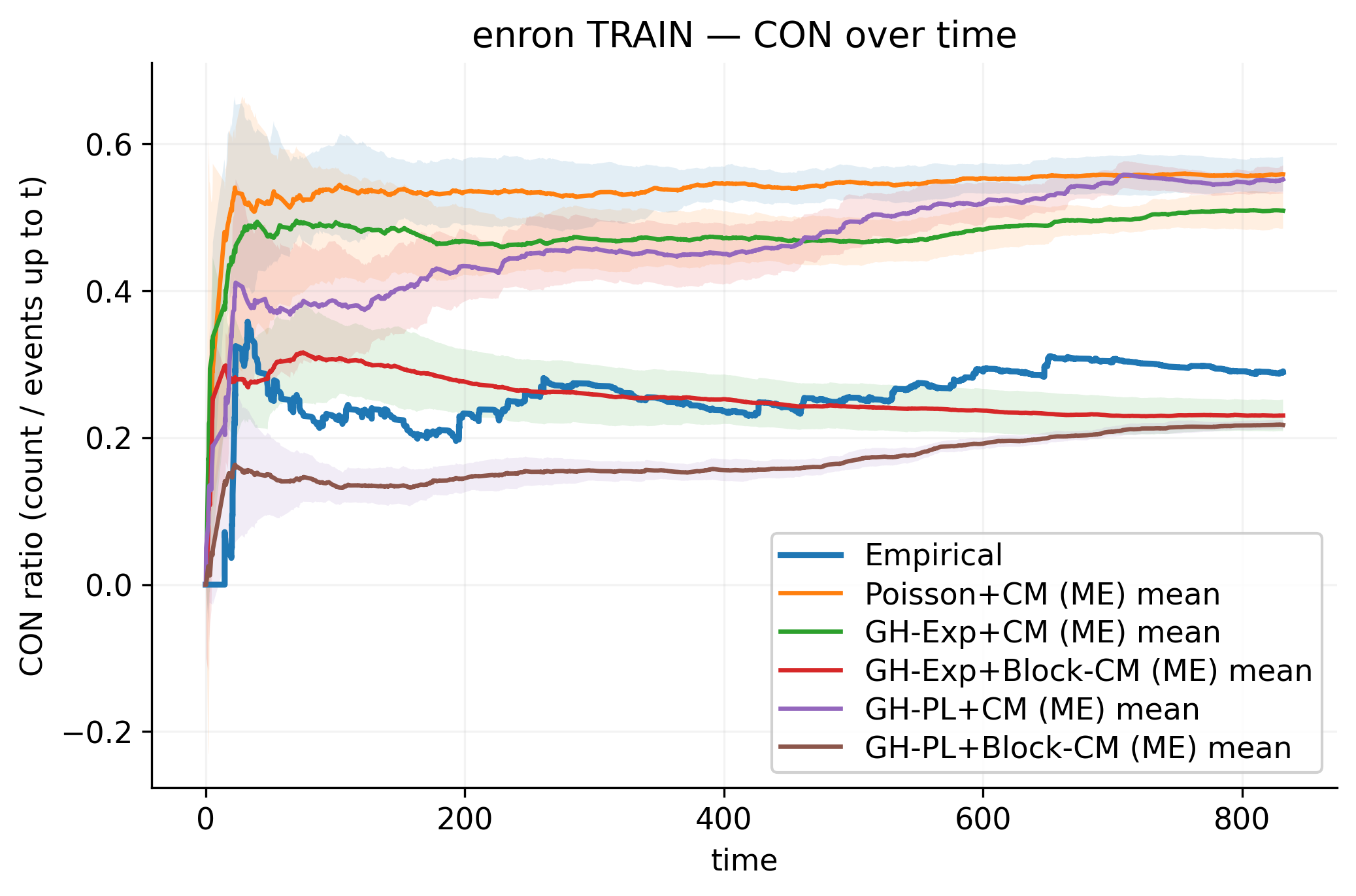}
    \caption{CON, $\Delta_{\text{large}}$}
  \end{subfigure}\hfill
  \begin{subfigure}{0.48\textwidth}
    \includegraphics[width=\linewidth]{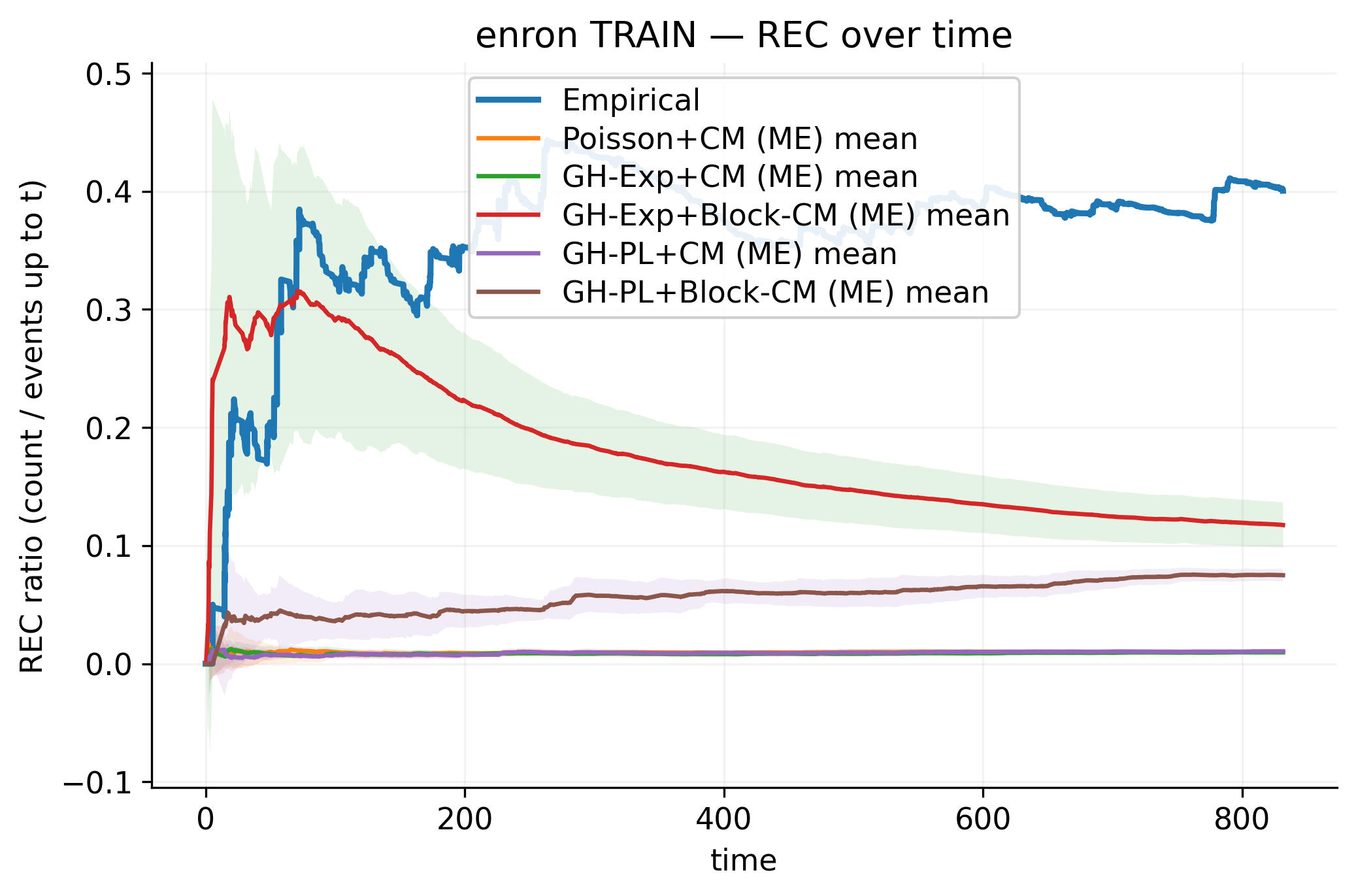}
    \caption{REC, $\Delta_{\text{large}}$}
  \end{subfigure}
  \vspace{0.75em}
  \begin{subfigure}{0.48\textwidth}
    \includegraphics[width=\linewidth]{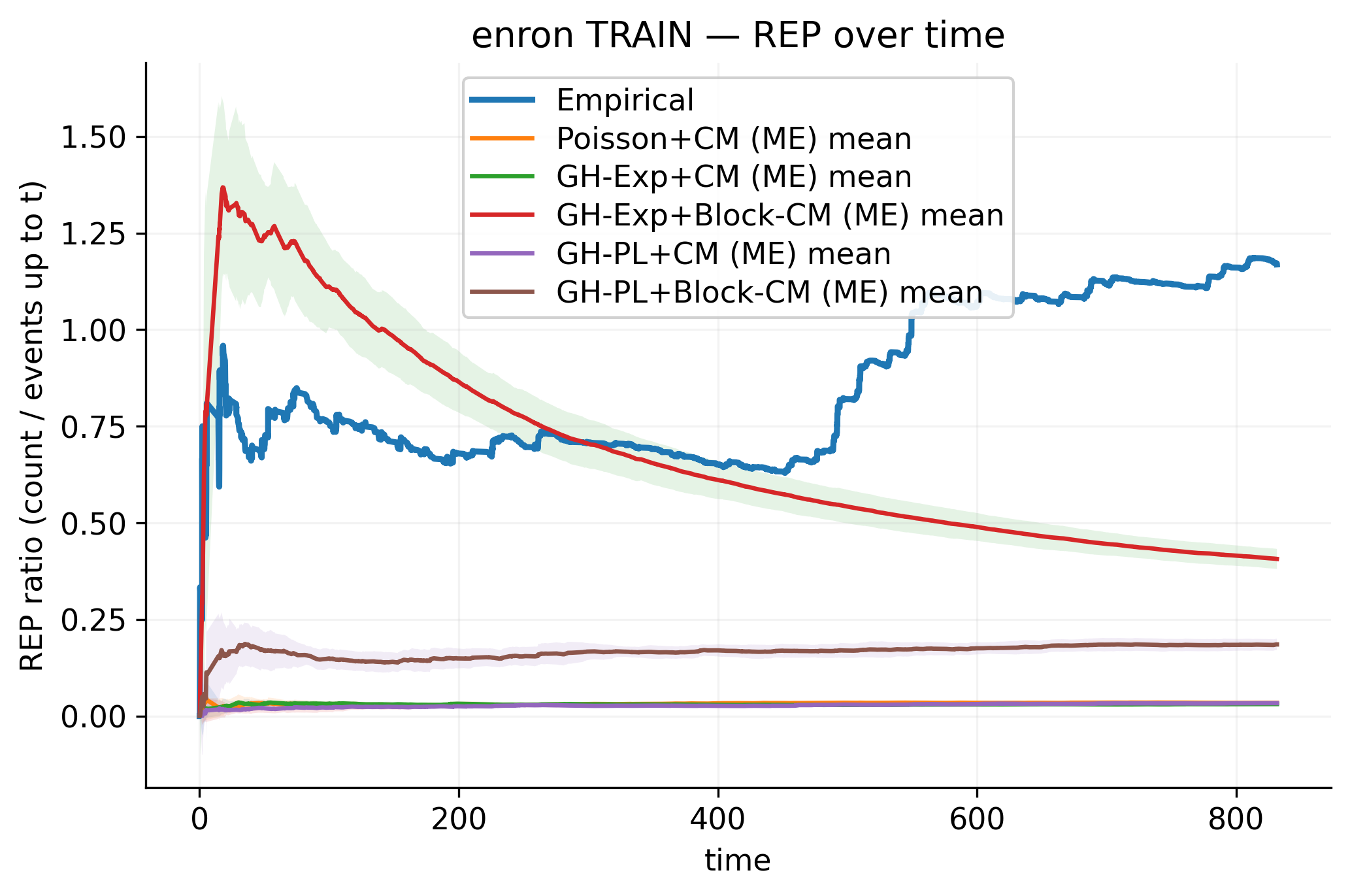}
    \caption{REP, $\Delta_{\text{large}}$}
  \end{subfigure}\hfill
  \begin{subfigure}{0.48\textwidth}
    \includegraphics[width=\linewidth]{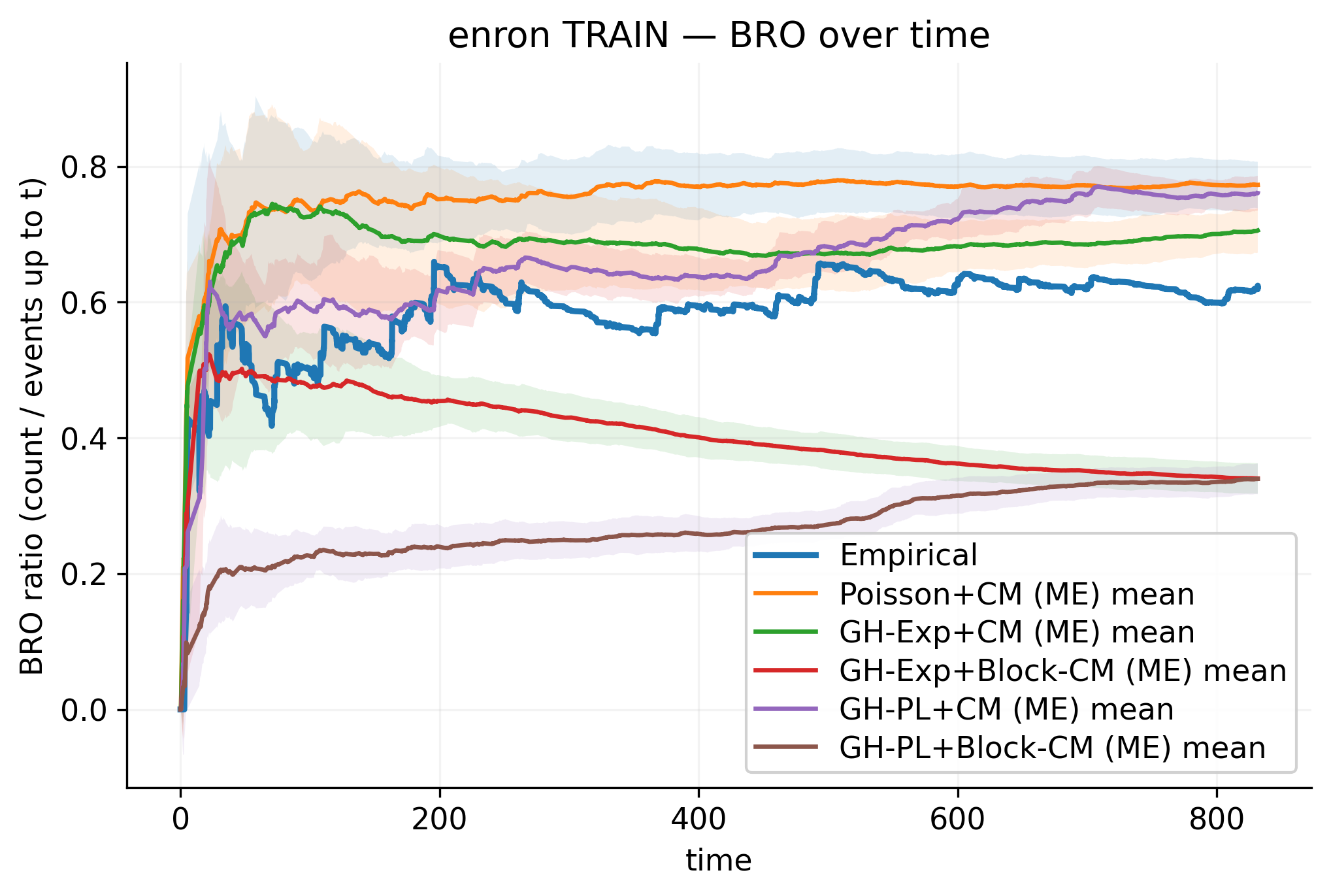}
    \caption{BRO, $\Delta_{\text{large}}$}
  \end{subfigure}
  \caption{Time-dependent motif ratios for the Enron dataset versus the maximum-entropy temporal network ensembles obtained combining Poisson/Exp-Hawkes (GH-Exp)/PL-Hawkes (GH-PL) temporal profiles and configuration model (CM) and block-configuration model (Block-CM) edge structures.}
  \label{fig:enron-motifs-large}
\end{figure*}

The methodology requires only event times and node pairs, and it is applicable to any temporal network dataset. 
The case study presented here serves as an example, but the same procedure can be extended to neuronal spike trains, epidemiological contact networks, and financial transactions.
In this sense, our framework provides a temporal-network analogue of static network maximum-entropy ensembles: a baseline against which higher-order structure can be detected, quantified, and interpreted across systems.

\section{Conclusion}
We have introduced a maximum-entropy framework for temporal networks in continuous time, combining non-homogeneous Poisson processes with static maximum-entropy edge distributions. The key consequence of the derivation is a clean factorization of time and marks, which yields closed-form likelihoods, tractable fitting procedures, and explicit formulas for edge coverage, burstiness, and motif statistics. 
The framework provides a benchmark for assessing which features are explained by basic structural and temporal constraints and which require additional mechanisms. In particular, the use of NHPP improves agreement with empirical inter-event distributions and motif counts, while the static maximum-entropy edge modeling accounts for degree and block structure. The resulting ensembles thus serve as principled null models for temporal networks, offering a consistent basis for significance testing and model comparison.
The present results allow immediate generalizations in several directions. The use of non-homogeneous Poisson processes enables us to fit an observed bursty behavior in the data or to impose it altogether with a theoretical model. In the maximum-entropy construction here presented there is no direct self-exciting behavior in the dataset generation, i.e. the self-excitation is driven by the original timestamped process. 
The introduction of this ensemble connecting  temporal networks to non-homogeneous Poisson processes provides a framework to naturally explore richer time profiles in networks: accounting for burstiness-degree correlations by introducing network-dependent factors in the intensity rates, including explicit heterogeneous calibration of the inter-event time distribution based on renewal processes theory, fitting block- and node-specific multivariate Hawkes processes, and generalising the kernel estimation to neural network functions \cite{mei2017neural}, thus contributing to bridging the gap between maximum entropy network ensembles and temporal graph-neural networks \cite{tian2025cohn}. 
The framework can be used to evaluate higher-order structures beyond two-event motifs, providing a systematic baseline for studying temporal clustering, cascades, or community dynamics. In all these cases, the maximum-entropy perspective clarifies what can be expected from constraints alone and where genuinely new patterns emerge in networked systems changing continuously over time.

\section*{Data and Code Availability}
All code used to generate the figures and reproduce the experiments is publicly available on \href{https://github.com/paolobarucca/Maximum-Entropy-Temporal-Networks}{github/Maximum-Entropy-Temporal-Networks} and \href{https://doi.org/10.5281/zenodo.18613086}{Zenodo}.
All figures are fully reproducible from the repository.

\bibliographystyle{apsrev4-2}
\bibliography{temporalnets}

\onecolumngrid

\section{Supplementary}
\section{Marked point processes, likelihood, and MaxEnt}
\label{sec:theory}

\paragraph*{Events and marks.} 
Let $V=\{1,\dots,n\}$ and $E\subseteq\{(i,j)\in V^2: i\neq j\}$ be the set of admissible directed edges. A temporal network is a marked point process on $[0,T]$ with events $\{(t_k,m_k)\}_{k=1}^K$, where $t_k\in(0,T]$ and $m_k=(i_k\!\to\!j_k)\in E$. We denote by $N_{ij}(I)$ the count of $(i\!\to\!j)$ events whose times lie in an interval $I\subseteq[0,T]$, and by $K(I)=\sum_{(i,j)\in E}N_{ij}(I)$ the total event count.

\paragraph*{Factorized intensity (time $\times$ marks).}
We fix a partition $P=\{E_r\}_{r=1}^R$ of $E$. The model intensity factorizes as
\begin{equation}
  \lambda_{ij}(t)=\phi_{r(i,j)}(t)\,w_{ij}, 
  \qquad (i,j)\in E,
  \label{eq:fact-basic}
\end{equation}
where: (i) $\phi_r(t)\ge 0$ is the \emph{time layer} for part $r$; (ii) $w_{ij}\ge 0$ are \emph{time–independent} mark weights (with $w_{ii}=0$). Define the per–part mass and integral
\begin{equation}
  W_r := \sum_{(i,j)\in E_r} w_{ij},
  \qquad
  \Lambda_r(I):=\int_I \phi_r(t)\,W_r\,\dd t,
\end{equation}
and the normalized mark law in part $r$,
\begin{equation}
  \Pi^{(r)}_{ij}
  := \begin{cases}
    \dfrac{w_{ij}}{W_r}, & (i,j)\in E_r,\\[6pt]
    0,& \text{otherwise}.
  \end{cases}
  \qquad
  \sum_{(i,j)} \Pi^{(r)}_{ij}=1.
  \label{eq:pir}
\end{equation}
Then \eqref{eq:fact-basic} can be re-written as: 
\begin{equation}
  \lambda_{ij}(t)
  = \Pi^{(r)}_{ij}\,\lambda_r(t),
  \qquad \lambda_r(t):=\phi_r(t)\,W_r.
  \label{eq:mix}
\end{equation}

\paragraph*{Likelihood and exact split.}
For a marked non-homogeneous Poisson process, the log–likelihood is
\begin{equation}
  \ell(\{\lambda_{ij}\})
  = \sum_{k=1}^K \log \lambda_{i_k j_k}(t_k)
    - \sum_{(i,j)\in E}\int_0^T \lambda_{ij}(t)\,\dd t.
  \label{eq:LL}
\end{equation}
If \eqref{eq:fact-basic} holds globally ($R=1$, write $\phi,\;w$), then
\begin{equation}
  \ell(\phi,w)
  = \sum_{k=1}^K \log \phi(t_k) - W\!\int_0^T \phi(t)\,\dd t
    + \sum_{(i,j)} N_{ij}(I) \log w_{ij},
  \label{eq:LL-split-global}
\end{equation}
which separates \emph{time} (first two terms) and \emph{marks} (last term). For a general partition, the split reads
\begin{equation}
  \ell(\{\phi_r\},w)
  = \sum_{r=1}^R \Big(\sum_{k:\,m_k\in E_r}\log \phi_r(t_k) - W_r\!\int_0^T\phi_r(t)\,\dd t\Big)
    + \sum_{(i,j)} N_{ij}(I)\log w_{ij}.
  \label{eq:LL-split-part}
\end{equation}

\paragraph*{Maximum-entropy derivation and factorization theorem.}
Maximizing path entropy subject to linear constraints that (i) fix partition totals in time,
\begin{equation}
  \sum_{(i,j)\in E_r}\lambda_{ij}(t) = G_r(t)\quad\forall r,\;t,
  \label{eq:time-constraint}
\end{equation}
and (ii) fix time–integrated margins (expected edge totals $\mu_{ij}$, node strengths $s_i^{\text{out}}=\sum_jN_{ij}$, $s_j^{\text{in}}=\sum_iN_{ij}$, block totals $N_{ab}$, etc.), yields the Euler–Lagrange stationarity condition that leads to the factorization
\begin{equation}
  \lambda_{ij}(t) 
  = \exp\!\Big\{\alpha_{r(i,j)}(t) + \Theta_{ij} + \theta_i^{\rm out} + \theta_j^{\rm in} + \zeta_{g(i),g(j)} - 1\Big\}\!,
\end{equation}
i.e.\ the product form \eqref{eq:fact-basic} with $\phi_r(t)=e^{\alpha_r(t)}$ and $w_{ij}=\exp(\Theta_{ij}+\theta_i^{\rm out}+\theta_j^{\rm in}+\zeta_{g(i),g(j)}-1)$. 
Thus the factorization is the \emph{unique} maximum-entropy solution for linear time and mark constraints.\footnote{Forbidden edges can be enforced by a mask $M_{ij}\in\{0,1\}$ by adding $\log M_{ij}$ to the mark potential.}

\subsection{Fitted marks layer}
Given $G_r$ (or a parametric fit for $\lambda_r$) and chosen margins, one solves for $\Pi$ via masked biproportional scaling (IPFP/Sinkhorn):
\begin{align}
  \text{Edge totals:}\quad & \int_0^T\lambda_{ij}(t)\,\dd t=\mu_{ij} 
  \;\;\Rightarrow\;\; \Pi_{ij} = \mu_{ij}/\Lambda^{tot}.
\\
  \text{Strengths:}\quad
  & \sum_{j\neq i} \Pi_{ij} = \frac{s_i^{\rm out}}{\Lambda^{tot}} ,
  \qquad
    \sum_{i\neq j} \Pi_{ij} = \frac{s_j^{\rm in}}{\Lambda^{tot}},
\end{align}
where we are allocating the events generated by the time profiles $\lambda_r(t)$ across the available edges according to the matrix $\Pi$, and where $\Lambda^{tot}=\Lambda([0,T])$. This is equivalent to known maximum entropy models in static network ensembles \cite{cimini2019statistical}. 

\subsection{Fitted time layers }
\label{sec:time-layers}
For stationary linear Hawkes $\lambda_r(t)=\mu_r+\int_0^\infty g_r(u)\,\dd N_r(t-u)$ with branching ratio $\eta_r:=\int_0^\infty g_r(u)\dd u<1$, one has the stationary identities, \cite{hawkes1971point,bacry2015hawkes,daley2003point},
\begin{equation}
  \bar\lambda_r=\frac{\mu_r}{1-\eta_r},
  \qquad
  S_r(0)=\frac{\bar\lambda_r}{(1-\eta_r)^2},
  \qquad
  \bar\lambda_r + 2\!\int_0^\infty c_r(u)\,\dd u = S_r(0),
  \label{eq:hawkes-mom}
\end{equation}
where we introduced the Bartlett spectrum, $S_r(\omega)=\bar\lambda_r\big|1-\mathcal{G}_r(\omega)\big|^{-2}$, with $\mathcal{G}_r(\omega)=\int_0^\infty g_r(u)e^{-i\omega u}\dd u$ and for stationary linear Hawkes models, $c_r(u)$ is the inverse Fourier transform of the Bartlett spectrum. 
Main linear Hawkes cases:
\begin{align}
  \text{Poisson:}&\quad g_r\equiv 0 \Rightarrow \eta_r=0,\;\bar\lambda_r=\mu_r,\; S_r(0)=\bar\lambda_r,\;c_r\equiv 0.\\
  \text{Exp kernel:}&\quad g_r(u)=m_r\beta_r e^{-\beta_r u}\1_{u>0},\;\eta_r=m_r,\; \text{\eqref{eq:hawkes-mom} holds}.\\
  \text{Power-law kernel:}&\quad g_r(u)=\eta_r \frac{(\alpha_r-1)k_r^{\alpha_r-1}}{(u+k_r)^{\alpha_r}}\1_{u>0},\; \alpha_r>1,\; \text{\eqref{eq:hawkes-mom} holds if }\eta_r<1.
\end{align}
For a kernel $g_r(u)=\alpha_r e^{-\beta_r u}$ with branching ratio $\eta_r=\alpha_r/\beta_r<1$ and stationary rate $\bar\lambda_r=\mu_r/(1-\eta_r)$,
the intensity autocovariance decays exponentially,
\begin{equation}
  \gamma_r(u)
  = \bar\lambda_r\,\frac{\alpha_r^2}{2(\beta_r-\alpha_r)}\,
    e^{-(\beta_r-\alpha_r)|u|}.
  \label{eq:gamma-exp}
\end{equation}
In the maximum-entropy construction, once we fit the arbitrary point processes to the observed series of timestamps we use the corresponding intensities as a constraint for the inhomogeneous Poisson process, i.e. the timestamp generation is then history-independent in the maximum-entropy case. 

\section{Partition models}
\label{sec:hybrid-sender-receiver}

To combine structural and temporal heterogeneity, we couple
distinct \emph{mark models} (encoding edge--weight constraints)
with different \emph{temporal kernels}.
Each partition type uses a consistent mark allocation
but allows for different time dynamics.
Table~\ref{tab:mark-time-models} summarizes all model combinations.

\begin{table}[h!]
  \centering
  \caption{Combinations of mark and time models across partition types.}
  \label{tab:mark-time-models}
  \vspace{4pt}
  \begin{tabular}{lll}
    \toprule
    \textbf{Partition} 
      & \textbf{Mark (edge--weight) model} 
      & \textbf{Time model} \\
    \midrule
    Global
      & WCM (Weighted configuration model)
      & Poisson / Hawkes--Exp / Hawkes--PL \\[2pt]
    Block--pair
      & Blockwise WCM (block $\times$ block expected weights)
      & Poisson / Hawkes--Exp / Hawkes--PL \\[2pt]
    Static baseline
      & WCM
      & Homogeneous Poisson (no excitation) \\[2pt]
    \bottomrule
  \end{tabular}
\end{table}

\paragraph{Block--pair partition.}
Nodes are first assigned to structural communities or blocks
(e.g.\ via Louvain modularity or metadata labels).
Each ordered block pair $(a,b)$ forms a subpartition $r=(a,b)$
with its own temporal kernel $g_r(u)$
and mark matrix $\Pi^{(r)}_{ij}$ constrained by blockwise-WCM expectations.
This representation preserves block--to--block strength constraints
and allows distinct excitation dynamics across inter-- and intra--block activity.
Block pairs therefore capture mesoscale heterogeneity that is
missed by global or node--level partitions.

\subsection{Frozen–path NHPP surrogates}
\begin{figure*}[t]
  \centering
  \includegraphics[width=0.9\textwidth]{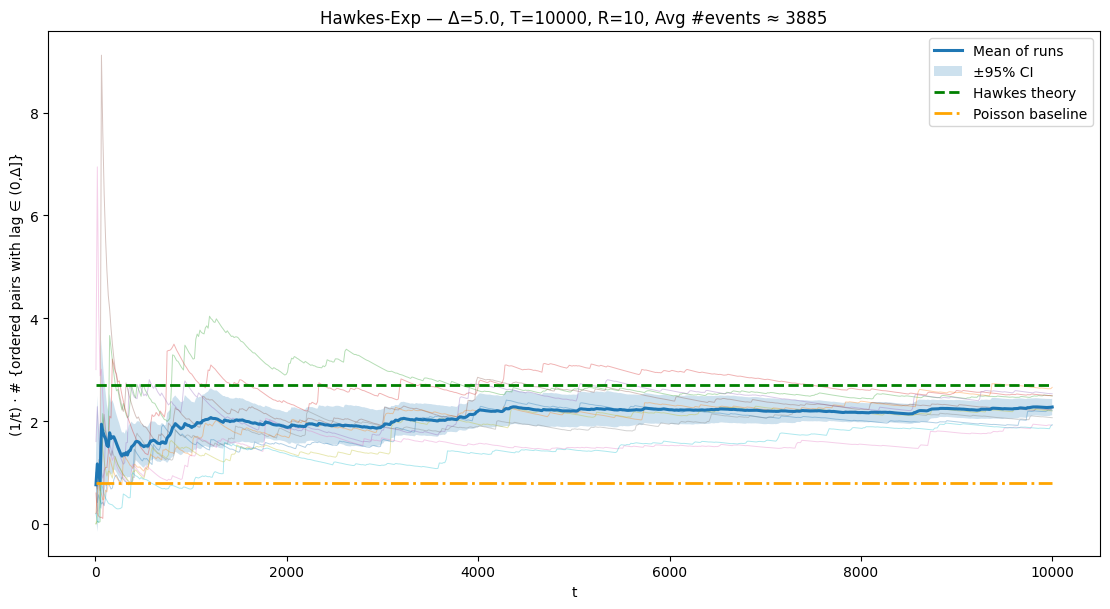}
  \caption{Frozen-path NHPP runs approximating the theoretical Hawkes auto-covariance.}
  \label{fig:frozen}
\end{figure*}

In the maximum-entropy construction, we are maximizing the entropy of a non-homogeneous Poisson process. Nevertheless, in many practical cases we would like to model the train of events with history-dependent stochastic processes, such as Hawkes process, doubly stochastic Poisson processes (Cox process), or renewal processes. 
Self–exciting processes such as the linear Hawkes model with exponential kernel, 
\begin{equation}
  \lambda(t)
  = \mu + \sum_{t_k<t} \alpha\,e^{-\beta(t-t_k)},
  \qquad
  \eta = \alpha/\beta < 1,
  \label{eq:hawkes-lambda}
\end{equation}
introduce \emph{stochastic feedback}: each event transiently raises the likelihood of future events.
The branching ratio $\eta$ controls the expected number of triggered descendants per event, and $\beta^{-1}$ sets the memory decay time.
In order to maintain the relevant properties of the stochastic models we can introduce \emph{frozen–path NHPP} surrogates, which preserve the fitted time–dependent rate $\lambda(t)$ of a stochastic model (e.g., Hawkes or renewal process) but remove stochastic feedback.
Formally, one replaces the random history-dependent intensity $\lambda(t|\mathcal{H}_t)$ by a deterministic path
$\hat{\lambda}(t)=\lambda(t|\mathcal{H}_t)$ and re-samples events as an NHPP with this frozen rate.
This surrogate maintains the expected counts $\Lambda(t)$, diurnal or seasonal modulations, and mark allocation structure, but suppresses the explicit stochastic feedback between events.
In general, for the individual frozen process for a given interval,
\begin{equation}
    \int_a^b\!\!\int_0^\Delta \hat{\lambda}(t)\hat{\lambda}(t+u)\,du\,dt
  \neq \int_a^b\!\!\int_0^\Delta \E[\lambda(t)\lambda(t+u)]\,du\,dt.
  \label{eq:frozen-expectation}
\end{equation}
However, when averaged over realizations of $\lambda(t)$ drawn from the original process, the expected pair–correlation of frozen NHPPs approximates that of the full process:
\begin{equation}
  \E_\lambda\!\left[
    \int_a^b\!\!\int_0^\Delta \hat{\lambda}(t)\hat{\lambda}(t+u)\,du\,dt
  \right]
  \sim \int_a^b\!\!\int_0^\Delta \E[\lambda(t)\lambda(t+u)]\,du\,dt,
  \label{eq:frozen-expectation}
\end{equation}
i.e. when averaging over realizations of the original point process the frozen-path autocovariance stays close to the theoretical stationary autocovariance of the stochastic point process, see Fig.\ref{fig:frozen}.
Thus, the frozen–path NHPP preserves the mean intensity and motif time factors in expectation, though any given realization lacks endogenous stochastic feedback.

\subsection{Inter–event times, rate variability, and burstiness}
\label{sec:isi}

For each edge $(i,j)$, let $\Delta_{ij}$ denote the \emph{inter–event time}, or inter–spike interval (ISI), i.e., the time difference between two consecutive events on that edge.  
For a stationary Poisson process with constant rate $\bar\lambda_{ij}$, events are independent and identically distributed, and inter–event times are exponentially distributed:
\begin{equation}
  p(\Delta_{ij}) = \bar\lambda_{ij}\, e^{-\bar\lambda_{ij}\Delta_{ij}},
  \qquad
  \E[\Delta_{ij}] = \frac{1}{\bar\lambda_{ij}},
  \qquad
  \sigma^2[\Delta_{ij}] = \frac{1}{\bar\lambda_{ij}^2}.
  \label{eq:isi-poisson}
\end{equation}
The process is memoryless, with coefficient of variation (CV) equal to one.
To quantify deviations from this baseline, define the coefficient of variation and the burstiness index:
\begin{equation}
  \mathrm{CV}_{ij}
  = \frac{\sqrt{\sigma^2[\Delta_{ij}]}}{\E[\Delta_{ij}]},
  \qquad
  B_{ij}
  = \frac{\mathrm{CV}_{ij}-1}{\mathrm{CV}_{ij}+1},
  \label{eq:burstiness}
\end{equation}
where $B_{ij}=0$ for a Poisson process, $B_{ij}>0$ for bursty dynamics (over–dispersion), and $B_{ij}<0$ for regular, quasi–periodic dynamics.

If the instantaneous rate $\lambda_{ij}(t)$ varies over time, events remain conditionally independent given $\lambda_{ij}(t)$, but the process is no longer stationary.
Periods of high activity can produce clusters of short inter–spike intervals, while low–rate periods generate long gaps.
Hence, a NHPP can exhibit \emph{apparent burstiness} ($B_{ij}>0$) arising entirely from deterministic rate modulation, even though events are conditionally Poisson and possess no genuine memory.
Thus, renewal and Hawkes models can generate identical mean rates but differ sharply in their conditional ISI structure.
Burstiness in event timing can therefore originate from two fundamentally distinct mechanisms:
\begin{itemize}
  \item exogenous modulation: deterministic variation in $\lambda(t)$ (as in a NHPP or frozen–path surrogate), which produces clustered intervals via rate heterogeneity but no true conditional dependence; and
  \item endogenous excitation: stochastic feedback (as in Hawkes or Cox processes), which generates genuine temporal correlations and amplifies variance even at fixed rate.
\end{itemize}
Both mechanisms yield broad inter–event–time distributions and $B_{ij}>0$, but only the latter reflects intrinsic memory in the event sequence.
The comparative analysis of homogeneous Poisson, time–varying NHPP, renewal, and Hawkes models thus can disentangle deterministic rate fluctuations from stochastic self–excitation as sources of bursty dynamics.

\section{Explicit expectations in time $\times$ marks form}
\label{sec:expectations}

All expectations below depend only on $(\lambda_r)$ and $(\Pi^{(r)})$ via \eqref{eq:mix}.

\subsection{Counts, strengths, rates}
For any interval $I=[a,b]\subseteq[0,T]$,
\begin{align}
  \E\big[N_{ij}(I)\big]
  &= \Pi^{(r)}_{ij}\; \Lambda_r(I), \,\, (i,j) \in E_r
  \label{eq:e-ij}
\\
  \E\big[K(I)\big] 
  &= \sum_{(i,j)} \E[N_{ij}(I)] = \sum_{r=1}^R \Lambda_r(I),
\\
  \E\big[s_i^{\rm out}(I)\big]
  &= \sum_j \E[N_{ij}(I)]
   = \sum_r \Lambda_r(I)\sum_j \Pi^{(r)}_{ij},
\\
  \E\big[s_j^{\rm in}(I)\big]
  &= \sum_i \E[N_{ij}(I)] 
   = \sum_r \Lambda_r(I) \sum_i \Pi^{(r)}_{ij}.
\end{align}
If each $\lambda_r$ is (wide–sense) stationary with mean $\bar\lambda_r$, the long–run edge rate is
\begin{equation}
  \bar\lambda_{ij} := \lim_{|I|\to\infty}\frac{\E [N_{ij}(I)]}{|I|}
  = \Pi^{(r)}_{ij}\,\bar\lambda_r.
  \label{eq:edge-rate}
\end{equation}

\subsection{Expected unique directed edges in $I$}
\label{sec:unique}
Under conditional Poisson arrivals, the per–edge count over $I$ is Poisson with mean $\mu_{ij}(I):=\E[N_{ij}(I)]$ from \eqref{eq:e-ij}. Thus the expected number of \emph{distinct} directed edges observed in $I$ is
\begin{equation}
  \E[U_{\to}(I)] 
  = \sum_{i\neq j}\big(1-e^{-\mu_{ij}(I)}\big),
  \qquad
  \mu_{ij}(I)
  = \Pi^{(r)}_{ij}\,\Lambda_r(I).
  \label{eq:unique}
\end{equation}
Block–pair unique–edge expectations follow by summation over $(i,j)$ with $g(i)=a,g(j)=b$.

\subsection{Motif pairs in a tolerance window $\Delta$}
\label{sec:motifs}
Consider ordered pairs of events with lag $u\in(0,\Delta]$. For two parts $r,s$ define sender/receiver masses
\begin{equation}
  r^{(r)}_i:=\sum_j \Pi^{(r)}_{ij},
  \qquad
  c^{(r)}_j:=\sum_i \Pi^{(r)}_{ij}.
\end{equation}
The \emph{pure–mark} probabilities, i.e.\ the chance that the second edge satisfies a motif given the first edge, are
\begin{align}
  p_{\textsc{rep}}^{(r,s)} &= \sum_{i,j} \Pi^{(r)}_{ij}\,\Pi^{(s)}_{ij}, 
  &
  p_{\textsc{rec}}^{(r,s)} &= \sum_{i,j} \Pi^{(r)}_{ij}\,\Pi^{(s)}_{ji}, 
  \label{eq:p-mark-1}
\\
  p_{\textsc{bcast}}^{(r,s)} &= \sum_i r^{(r)}_i\, r^{(s)}_i - p_{\textsc{rep}}^{(r,s)}, 
  &
  p_{\textsc{conv}}^{(r,s)} &= \sum_j c^{(r)}_j\, c^{(s)}_j - p_{\textsc{rep}}^{(r,s)}.
  \label{eq:p-mark-2}
\end{align}
The \emph{time} contribution is the per–part intensity cross–moment integrated over the window:
\begin{equation}
  C_{r,s}(\Delta;I)
  := \int_{a}^{b}\int_{0}^{\Delta} \E\big[\lambda_r(t)\lambda_s(t+u)\big]\,\dd u\,\dd t,
  \qquad I=[a,b].
  \label{eq:Crstime}
\end{equation}
Then the expected motif counts are
\begin{equation}
  \boxed{\;
  \E\big[M_{\mathrm{TYPE}}^{(\Delta)}(I)\big]
  = \sum_{r,s=1}^R p_{\mathrm{TYPE}}^{(r,s)}\, C_{r,s}(\Delta;I),
  \quad \mathrm{TYPE}\in\{\textsc{rec},\textsc{bcast},\textsc{conv},\textsc{rep}\}.
  \;}
  \label{eq:motif-master}
\end{equation}
Two explicit evaluations are useful:
\begin{itemize}
  \item If $r\neq s$ are independent layers, 
  $\E[\lambda_r(t)\lambda_s(t+u)]=\bar\lambda_r\,\bar\lambda_s$ and
  \begin{equation}
    C_{r,s}(\Delta;I)=|I|\,\Delta\,\bar\lambda_r\,\bar\lambda_s.
  \end{equation}
  \item For $r=s$, decompose 
  $\E[\lambda_r(t)\lambda_r(t+u)]=\bar\lambda_r^2+\gamma_r(u)$ to get
  \begin{equation}
    C_{r,r}(\Delta;I)=|I|\Big(\bar\lambda_r^2\,\Delta + \int_0^\Delta \gamma_r(u)\,\dd u\Big).
    \label{eq:Crr}
  \end{equation}
\end{itemize}
Integrating \eqref{eq:gamma-exp} in \eqref{eq:Crr} gives
\begin{equation}
  C_{r,r}(\Delta;I)
  = |I|\!\left[
    \bar\lambda_r^2\,\Delta
    + \bar\lambda_r\,\frac{\alpha_r^2}{2(\beta_r-\alpha_r)^2}
      \big(1 - e^{-(\beta_r-\alpha_r)\Delta}\big)
  \right].
  \label{eq:Crr-exp}
\end{equation}
The first term represents the Poisson baseline, while the second
quantifies the temporal amplification due to self–excitation,
which grows with the excitation strength $\alpha_r$ and the kernel memory $\beta_r^{-1}$.

\subsection{Model selection checks and identifiability}
\label{sec:ident}
\paragraph*{Scaling invariance.}
If $w\mapsto c\,w$ and $\phi_r\mapsto \phi_r/c$ for all $r$ (or $W_r\mapsto cW_r$ and $\phi_r\mapsto \phi_r/c$), then $\lambda_{ij}$ is unchanged. Any fitting scheme must fix this degree of freedom, e.g.\ by working with $(\lambda_r,\Pi^{(r)})$ directly, or by fixing $W_r$ (or $\int\phi_r$).

\paragraph*{Feasibility of margins.}
Strength/block totals must be compatible with the mask and per–part masses. A necessary condition is that each part saturates: $\sum_{(i,j)\in E_r} \mu_{ij}^\star=\Lambda_r([0,T])$; row/column sums must agree across decompositions. IPFP converges if and only if the mask admits a feasible point with those margins.

\paragraph*{Stability of Hawkes layers.}
Each Hawkes layer must satisfy $\eta_r<1$ to have finite $\bar\lambda_r$ and $S_r(0)$; otherwise expectations in \S\ref{sec:isi}–\S\ref{sec:motifs} are ill–posed.

\begin{figure*}[t]
  \centering
  \includegraphics[width=1\textwidth]{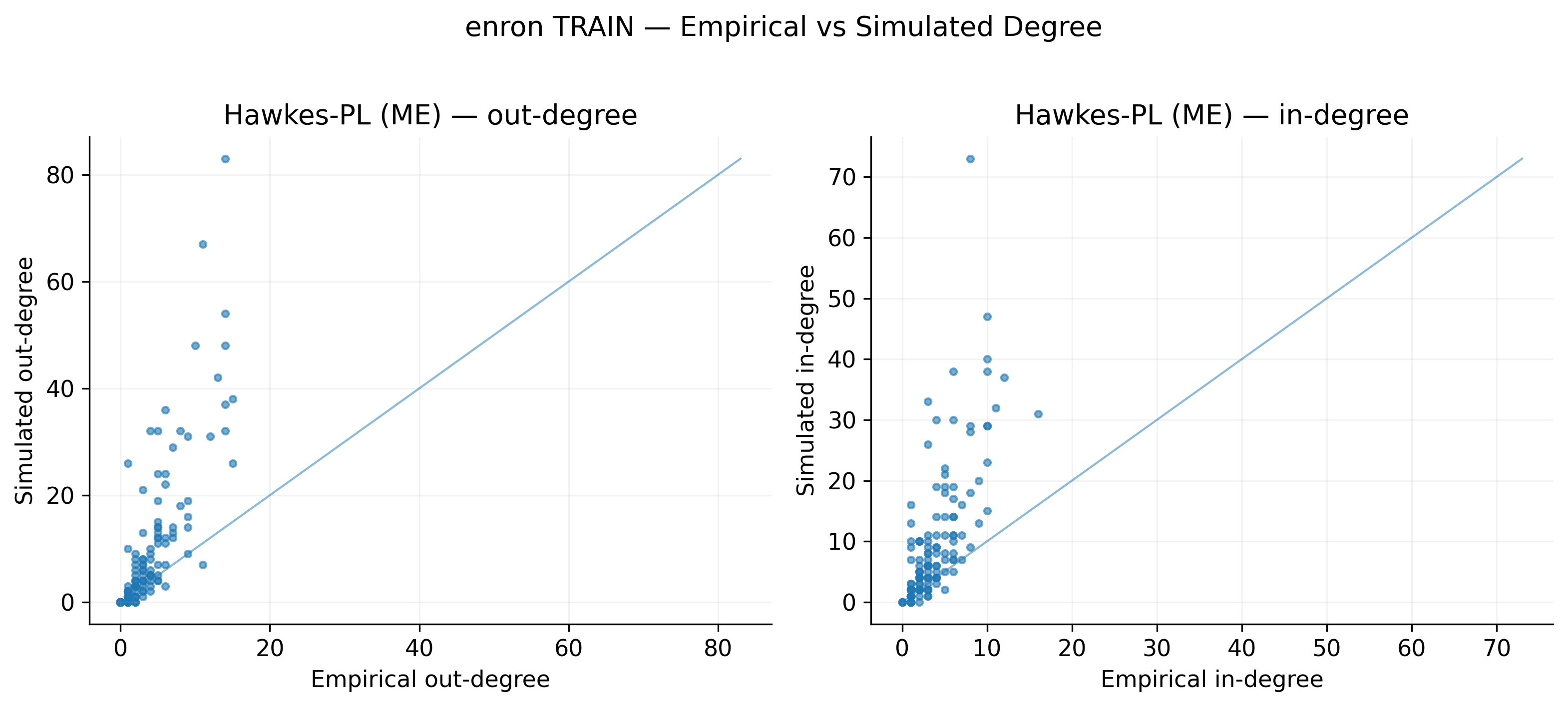}
  \caption{Degree-related statistics for Enron TRAIN (blockpair--PL model). 
  observed vs.\ expected out-degree}
  \label{fig:enron-degree}
\end{figure*}

\begin{figure*}[t]
  \centering
  \includegraphics[width=0.50\textwidth]{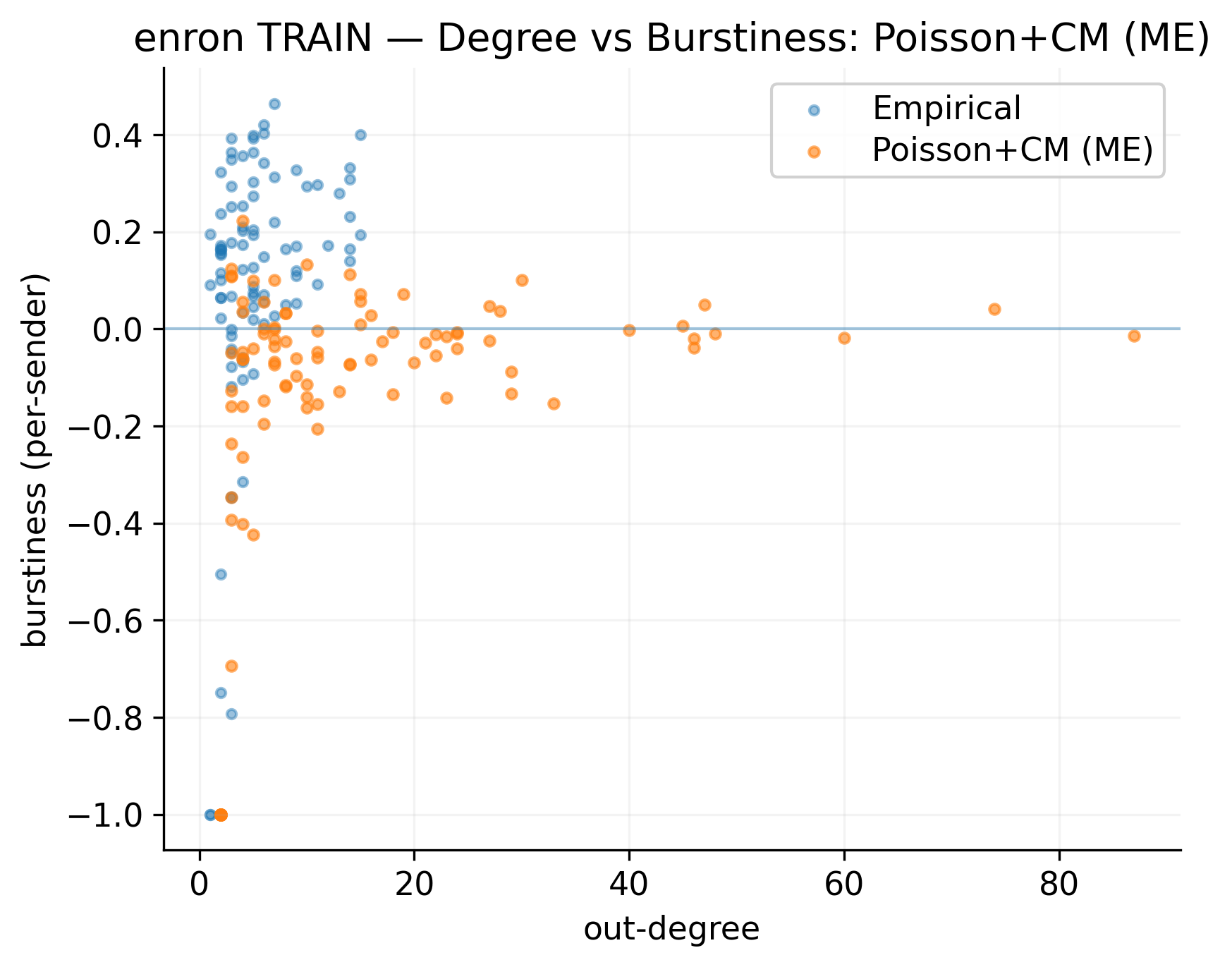}
  \caption{Burstiness vs.\ degree for the Enron TRAIN dataset.
  The plot compares empirical observations with the Poisson + Configuration Model, showing how burstiness scales with out-degree.}
  \label{fig:enron-clustering-vs-degree}
\end{figure*}

\begin{figure*}[t]
  \centering
  \includegraphics[width=1\textwidth]{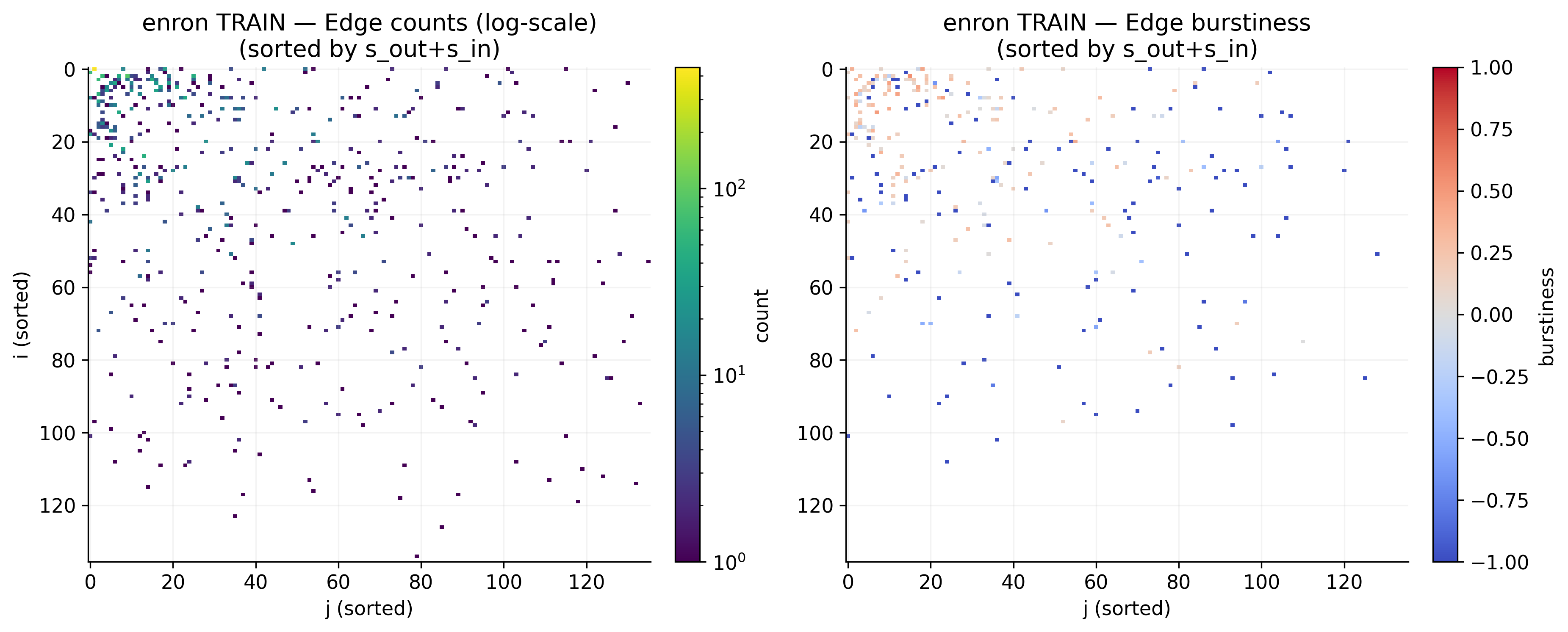}
  \caption{Log of event counts per edge (left) and Burstiness per edge (right) for the Enron TRAIN dataset.}
  \label{fig:enron-matrices}
\end{figure*}

\begin{figure*}[t]
  \centering
  \includegraphics[width=0.95\textwidth]{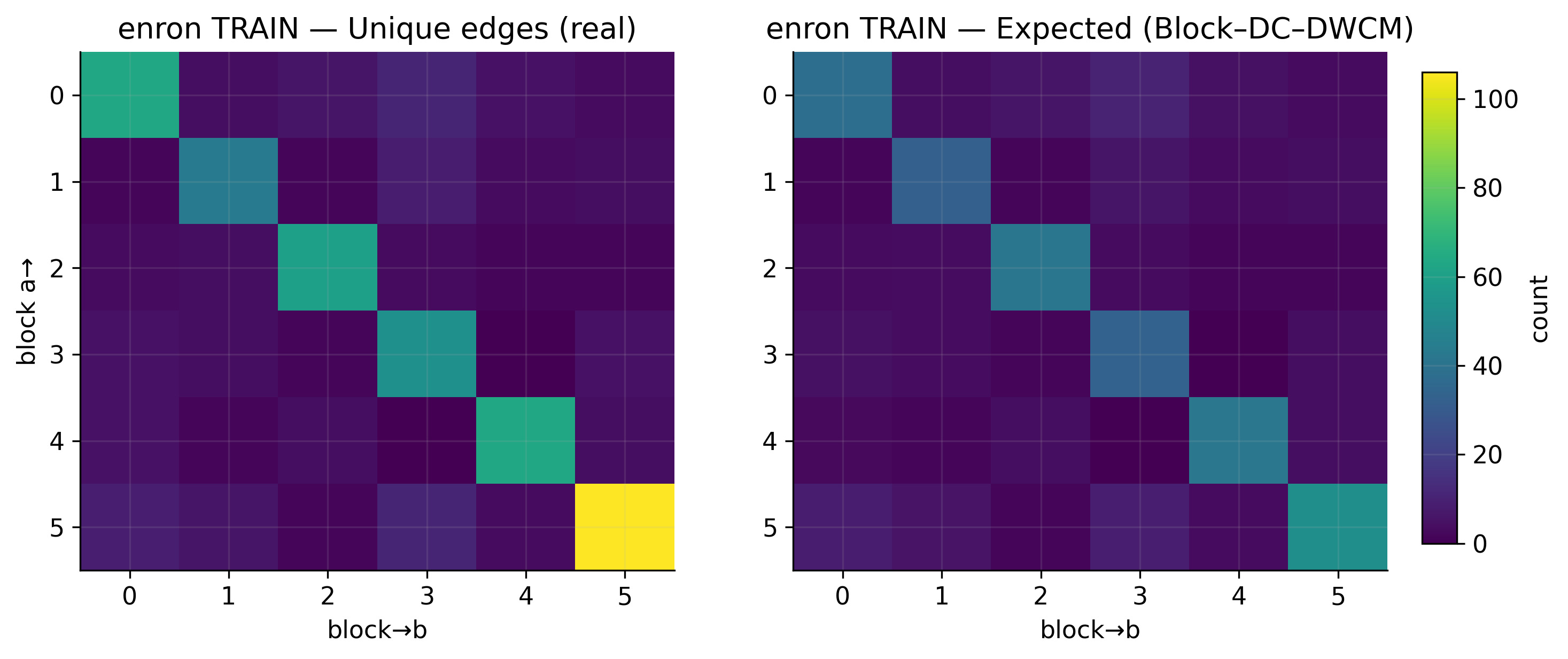}
  \caption{Block-to-block edge count comparison for Enron TRAIN:
  (\emph{left}) empirical unique edges, 
  (\emph{right}) expected under the Block–DC–DWCM constraint.}
  \label{fig:enron-blockpair-degree}
\end{figure*}

\begin{figure*}[t]
  \centering
  \includegraphics[width=0.75\textwidth]{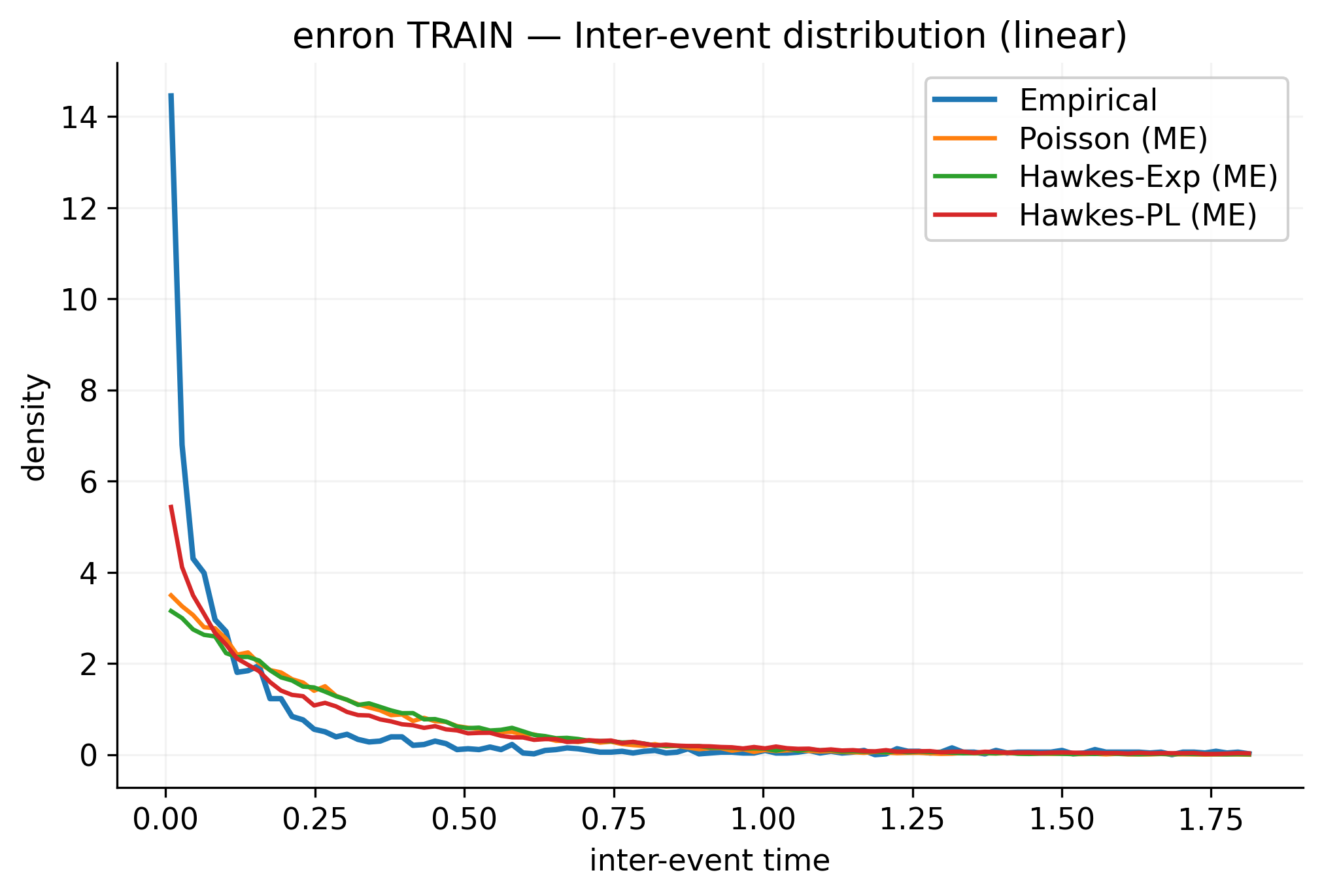}
  \caption{Inter-event time distributions (linear scale) for the Enron TRAIN dataset,
  comparing empirical data with fitted temporal models.
  The overlay highlights how each process reproduces or underestimates temporal burstiness.}
  \label{fig:enron-interevent-linear}
\end{figure*}

\begin{figure*}[t]
  \centering
  \includegraphics[width=\textwidth]{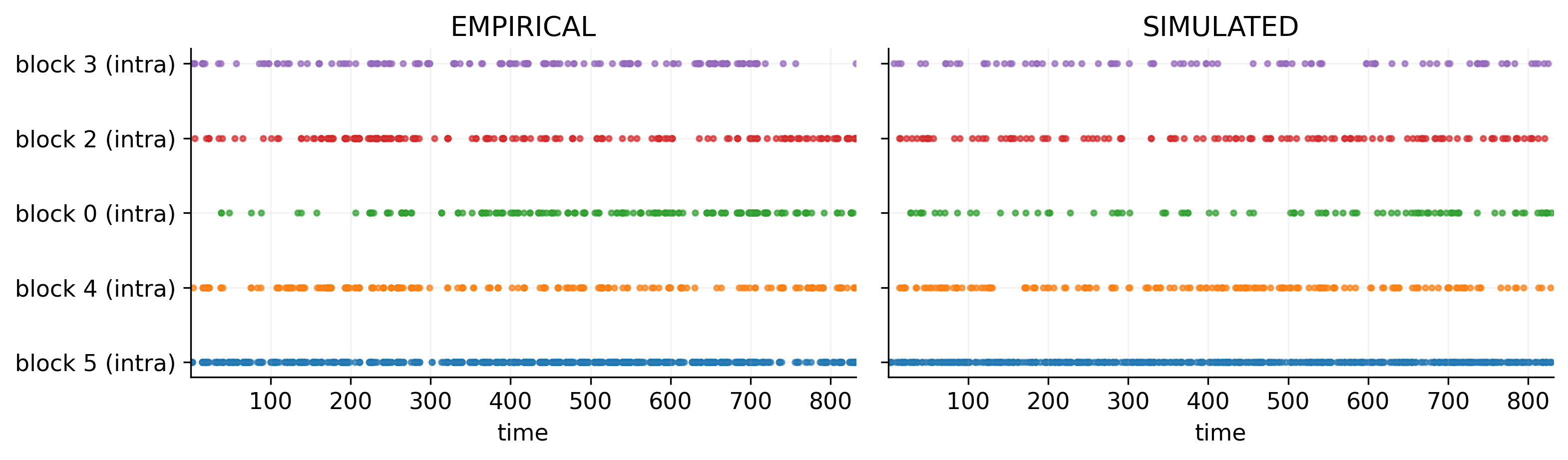}
  \caption{Raster plots of the top-5 intra-block event sequences for Enron TRAIN: 
  (\emph{left}) empirical; (\emph{right}) blockpair–PL model (sample $s{=}00$).}
  \label{fig:enron-raster-blocks}
\end{figure*}

\begin{figure*}[t]
  \centering
  \includegraphics[width=\textwidth]{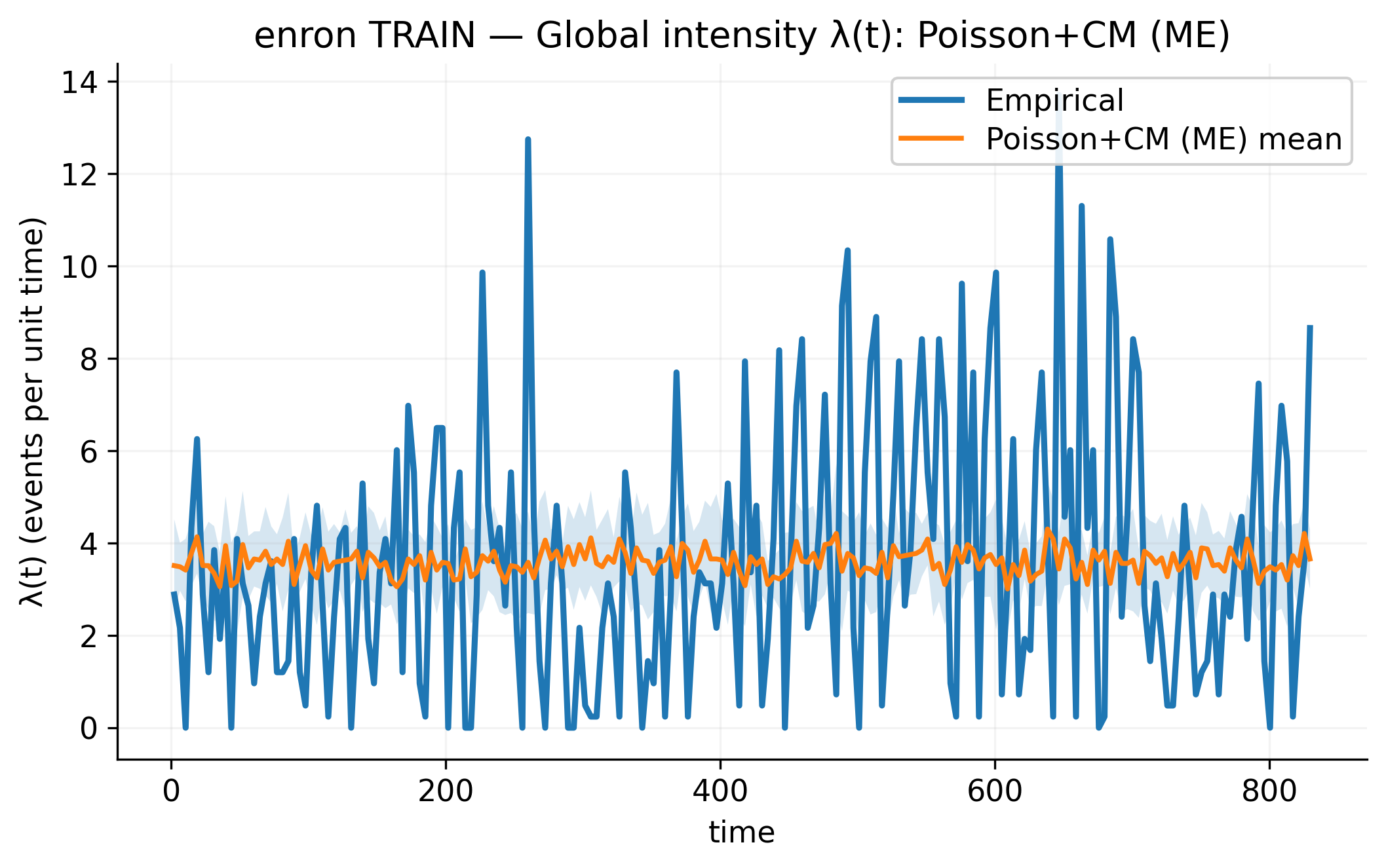}
\caption{Enron TRAIN intensity over time for the maximum-entropy Poisson + Configuration Model. }
  \label{fig:enron-lambda-Poisson}
\end{figure*}

\begin{figure*}[t]
  \centering
\includegraphics{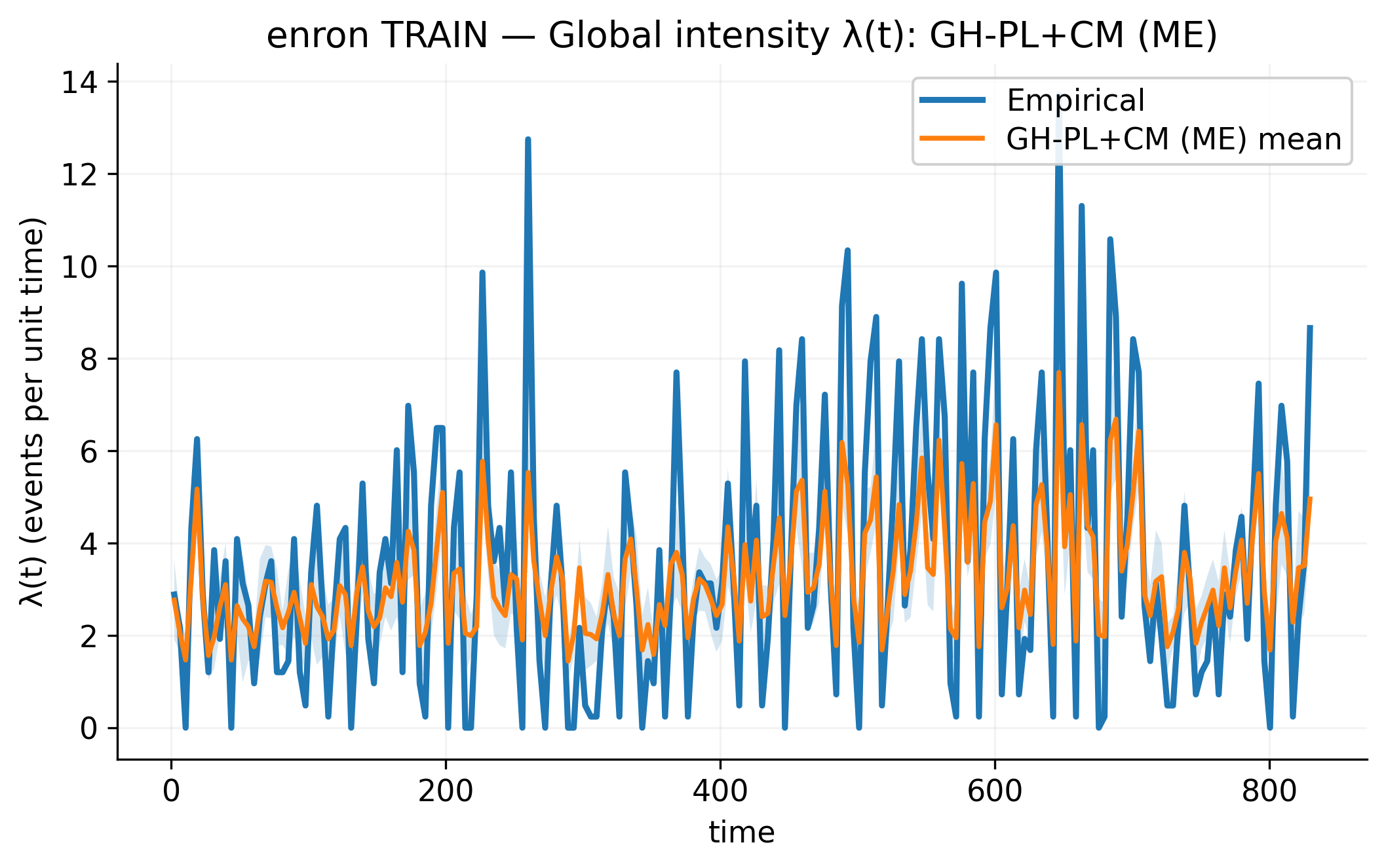}
\caption{Enron TRAIN intensity over time Enron TRAIN intensity over time for the maximum-entropy Global Power-Law Kernel Hawkes + Configuration Model.}
  \label{fig:enron-lambda-GH}
\end{figure*}

\subsection{Motif analysis across datasets}
\begin{figure*}[t]
  \centering
  \begin{subfigure}{0.48\textwidth}
    \includegraphics[width=\linewidth]{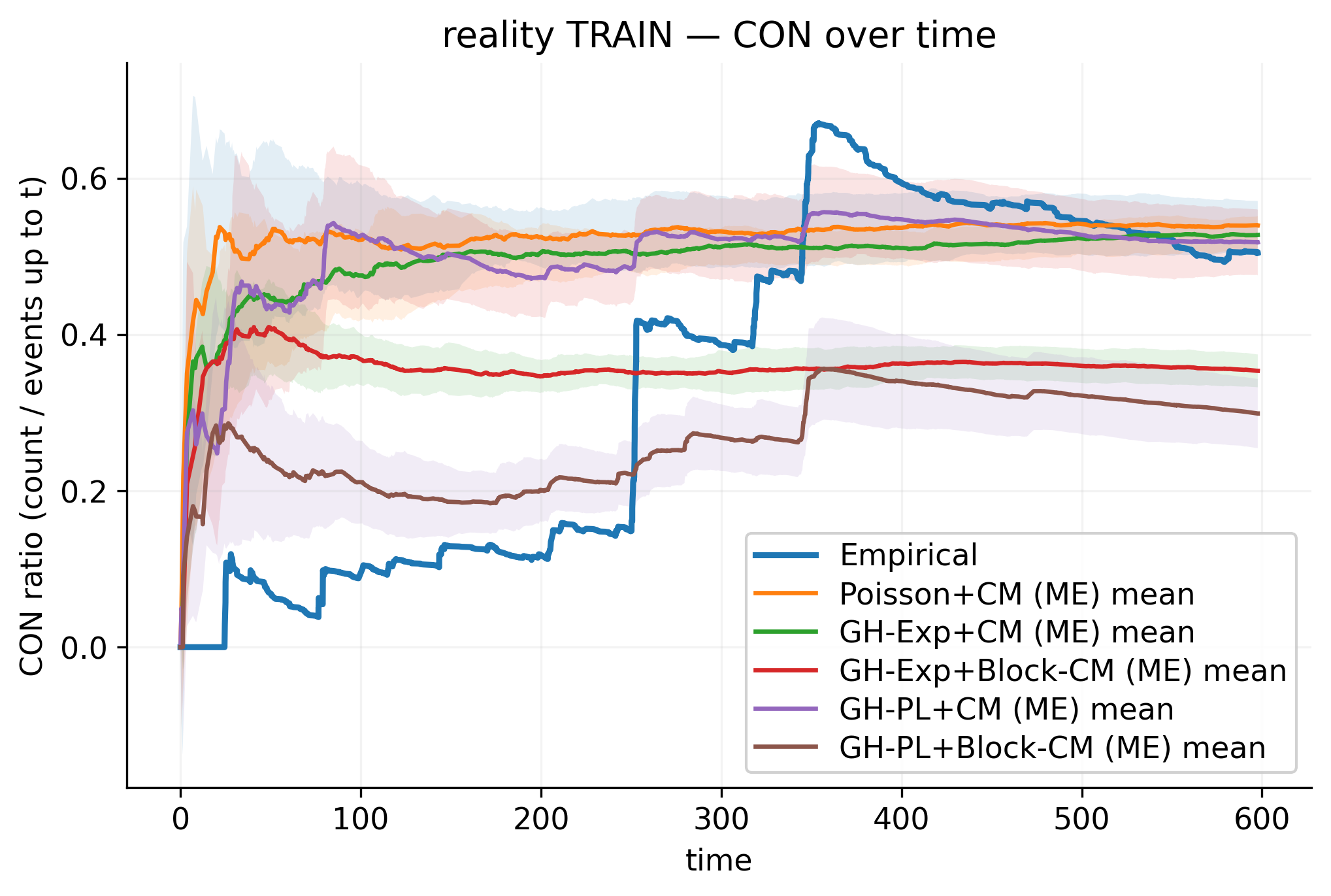}
    \caption{CON, $\Delta_{\text{large}}$}
  \end{subfigure}\hfill
  \begin{subfigure}{0.48\textwidth}
    \includegraphics[width=\linewidth]{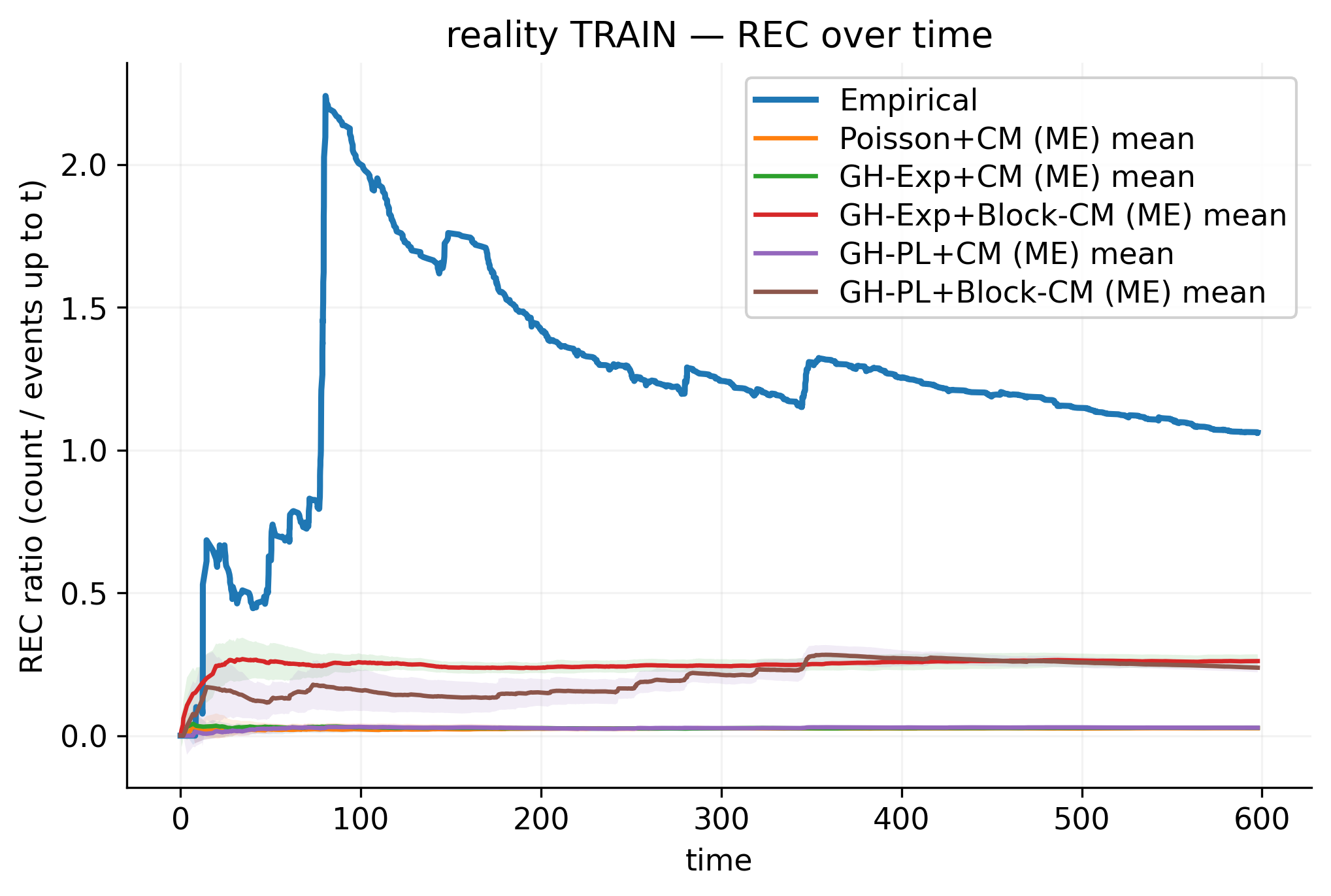}
    \caption{REC, $\Delta_{\text{large}}$}
  \end{subfigure}
  \vspace{0.75em}
  \begin{subfigure}{0.48\textwidth}
    \includegraphics[width=\linewidth]{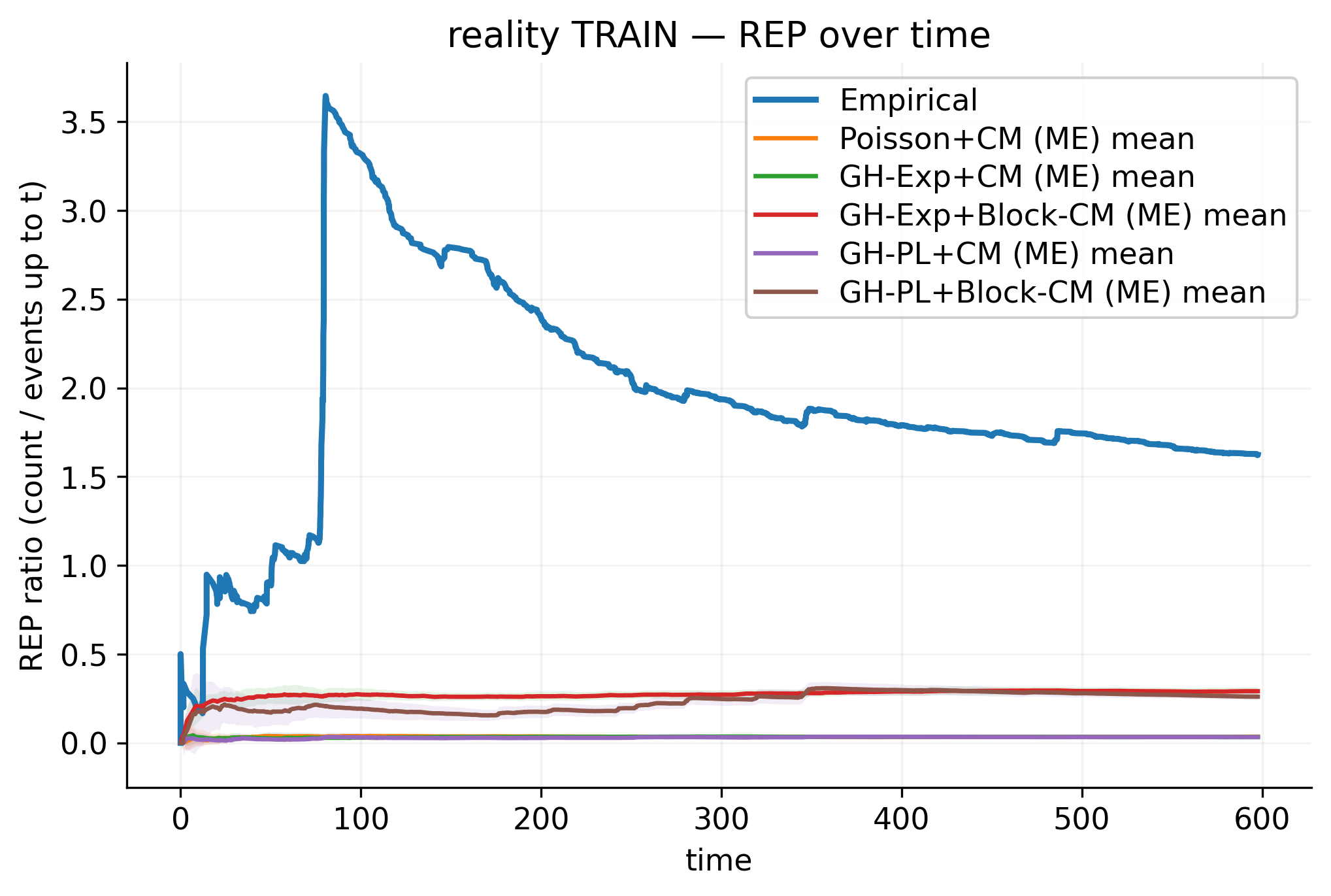}
    \caption{REP, $\Delta_{\text{large}}$}
  \end{subfigure}\hfill
  \begin{subfigure}{0.48\textwidth}
    \includegraphics[width=\linewidth]{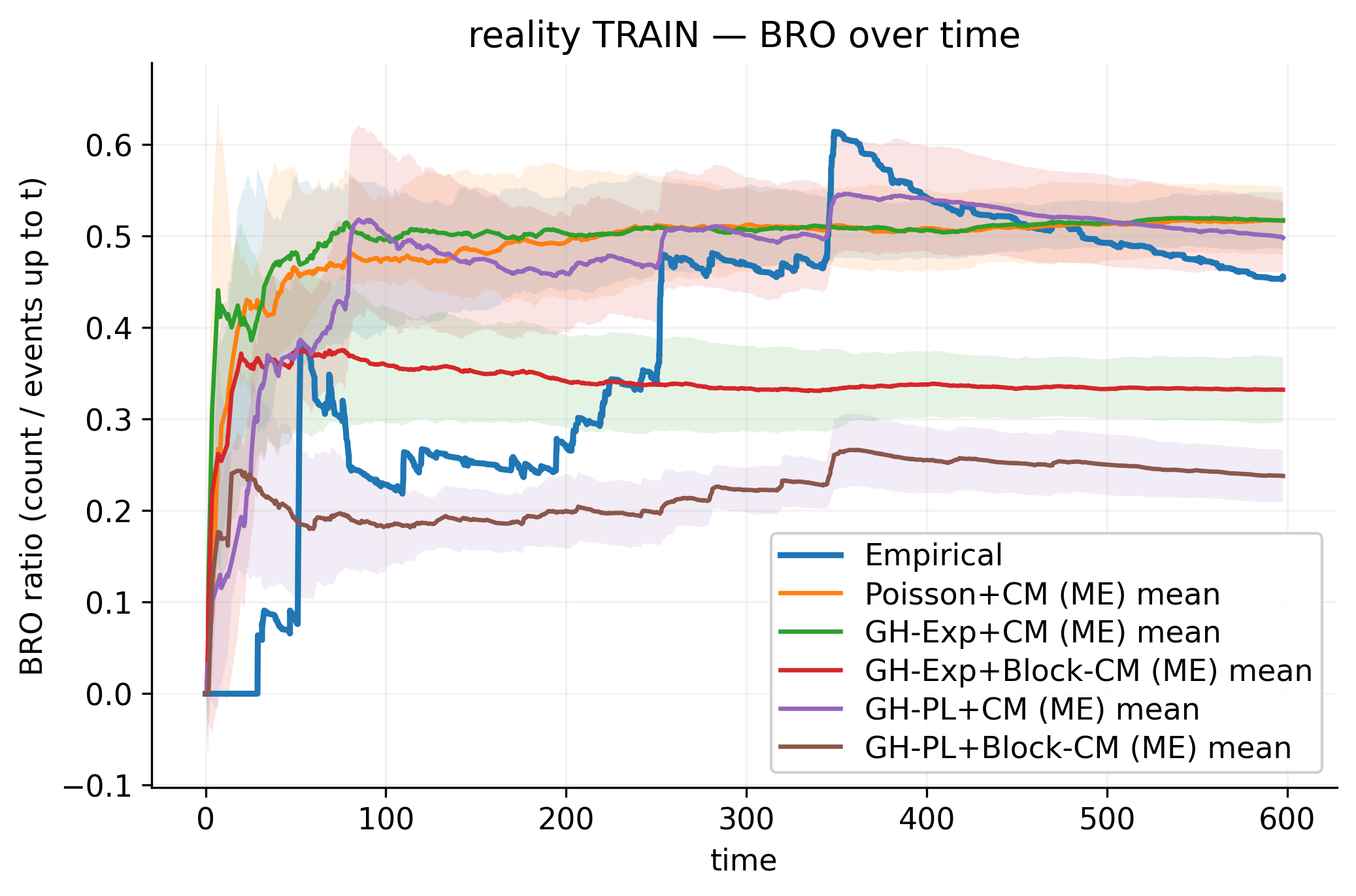}
    \caption{BRO, $\Delta_{\text{large}}$}
  \end{subfigure}
  \caption{Reality mining TRAIN motif ratios (mean $\pm$ s.d.) over time for $\Delta_{\text{large}}$.}
  \label{fig:reality-motifs-large}
\end{figure*}

\begin{figure*}[t]
  \centering
  \begin{subfigure}{0.48\textwidth}
    \includegraphics[width=\linewidth]{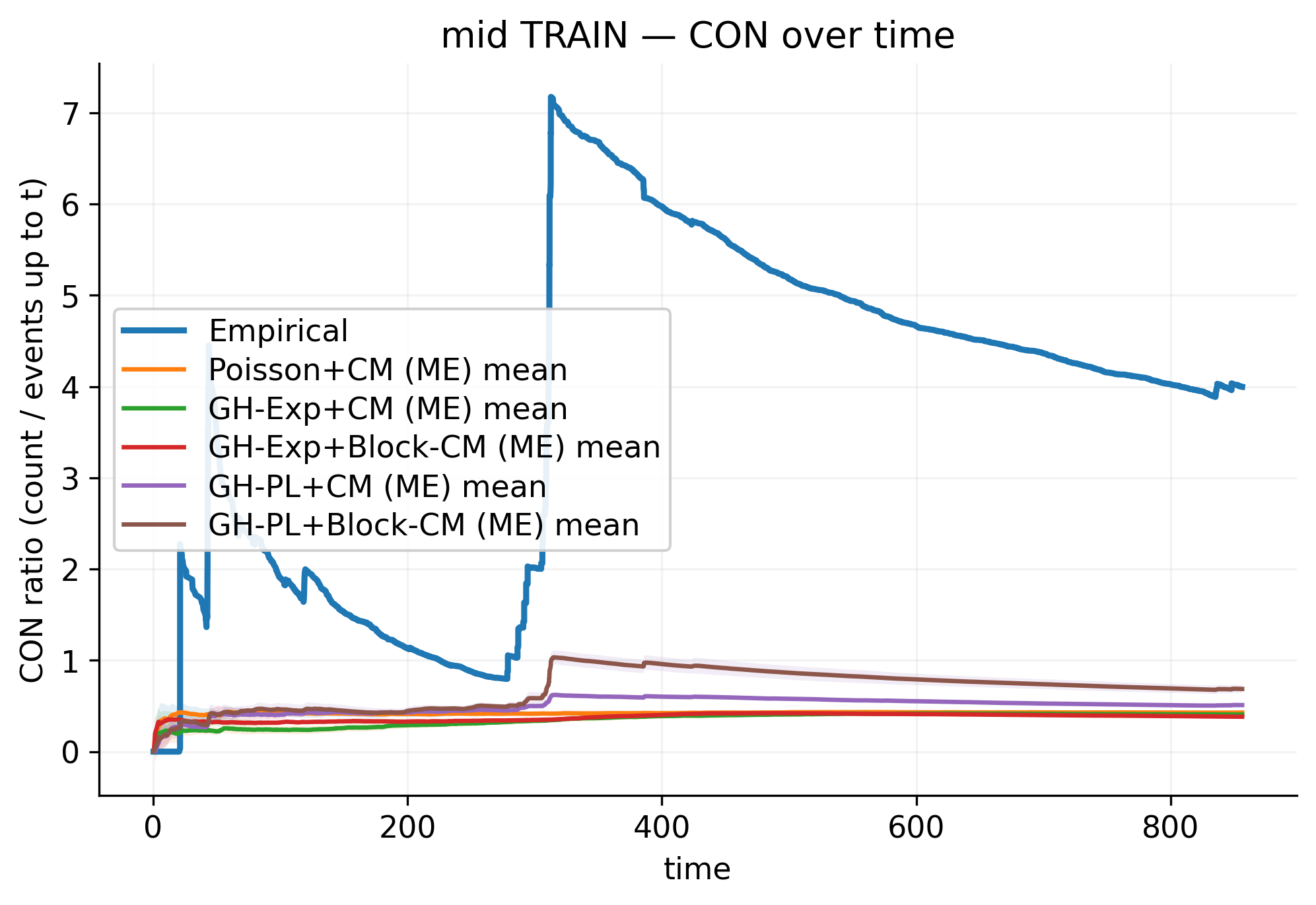}
    \caption{CON, $\Delta_{\text{large}}$}
  \end{subfigure}\hfill
  \begin{subfigure}{0.48\textwidth}
    \includegraphics[width=\linewidth]{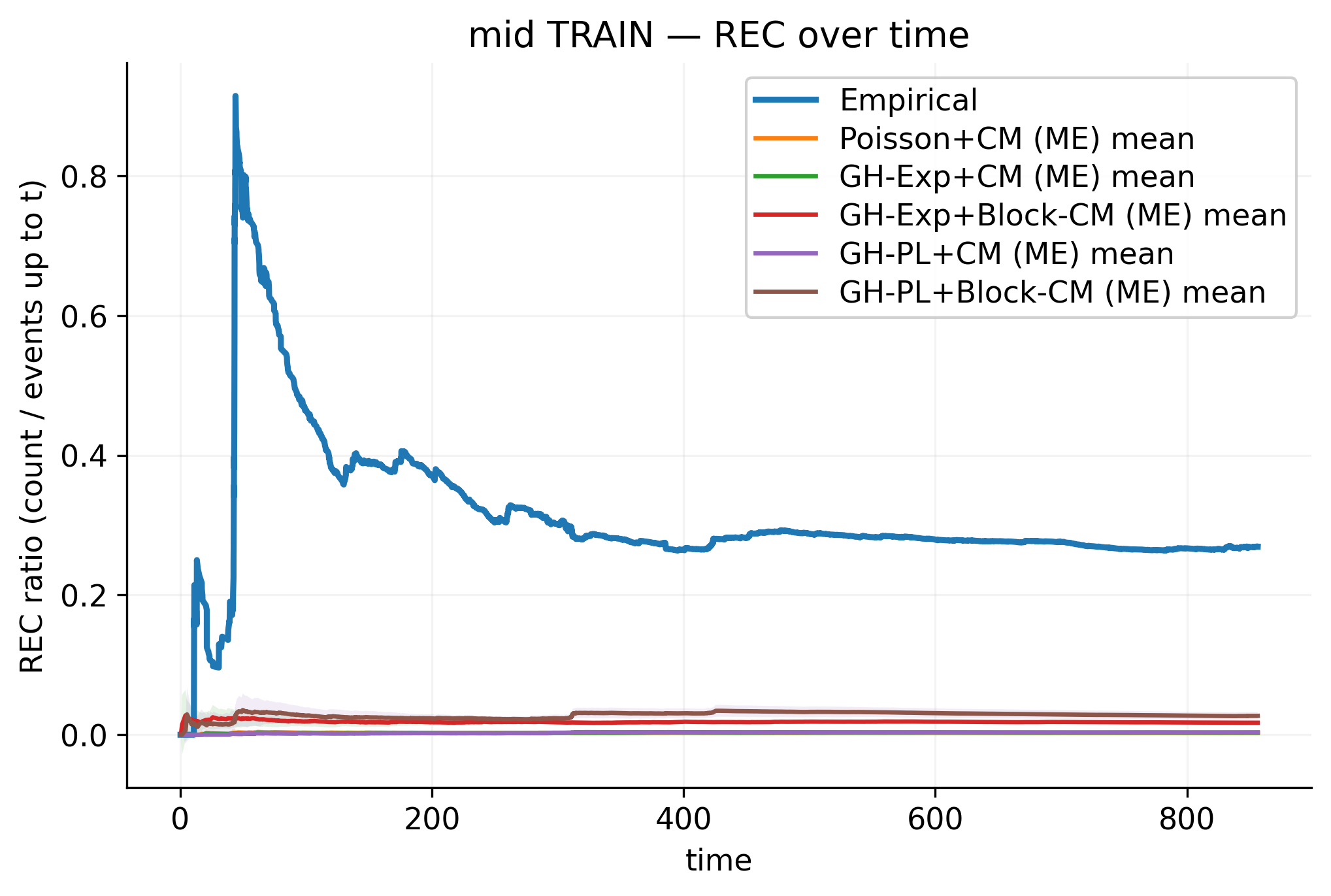}
    \caption{REC, $\Delta_{\text{large}}$}
  \end{subfigure}
  \vspace{0.75em}
  \begin{subfigure}{0.48\textwidth}
    \includegraphics[width=\linewidth]{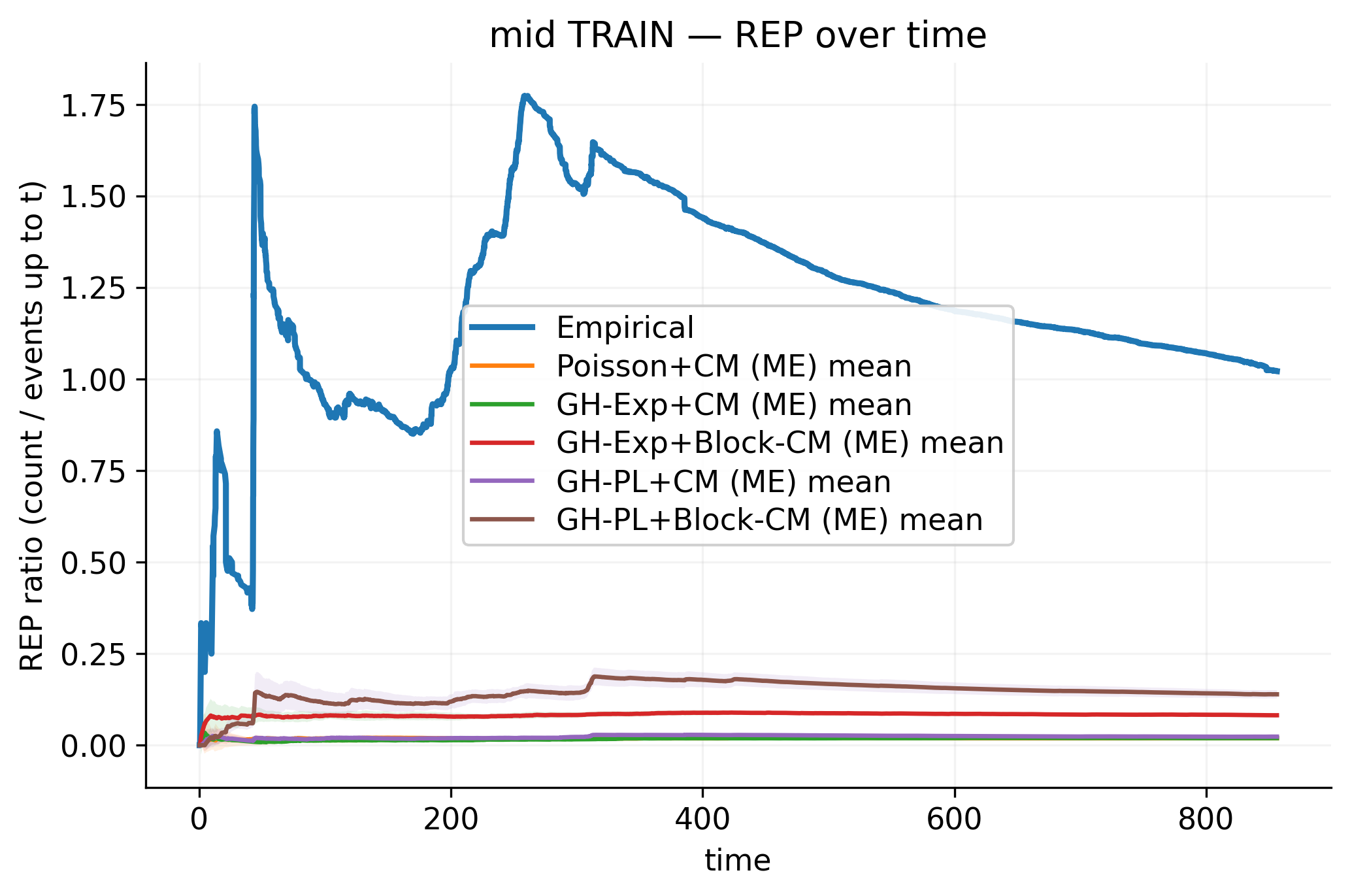}
    \caption{REP, $\Delta_{\text{large}}$}
  \end{subfigure}\hfill
  \begin{subfigure}{0.48\textwidth}
    \includegraphics[width=\linewidth]{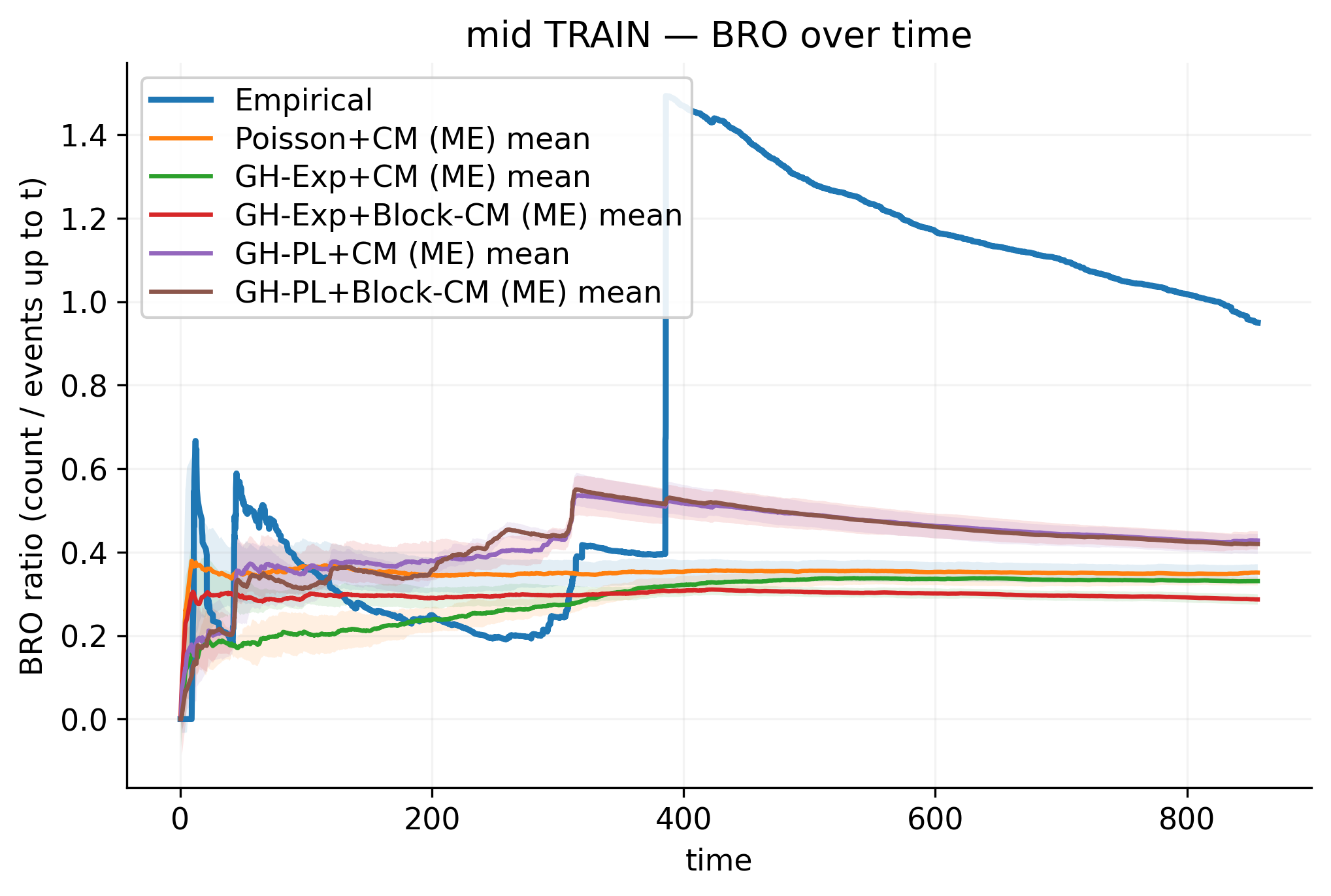}
    \caption{BRO, $\Delta_{\text{large}}$}
  \end{subfigure}
  \caption{Militarized Interstate Disputes  TRAIN motif ratios (mean $\pm$ s.d.) over time for $\Delta_{\text{large}}$.}
  \label{fig:mid-motifs-large}
\end{figure*}

\begin{figure*}[t]
  \centering
  \begin{subfigure}{0.48\textwidth}
    \includegraphics[width=\linewidth]{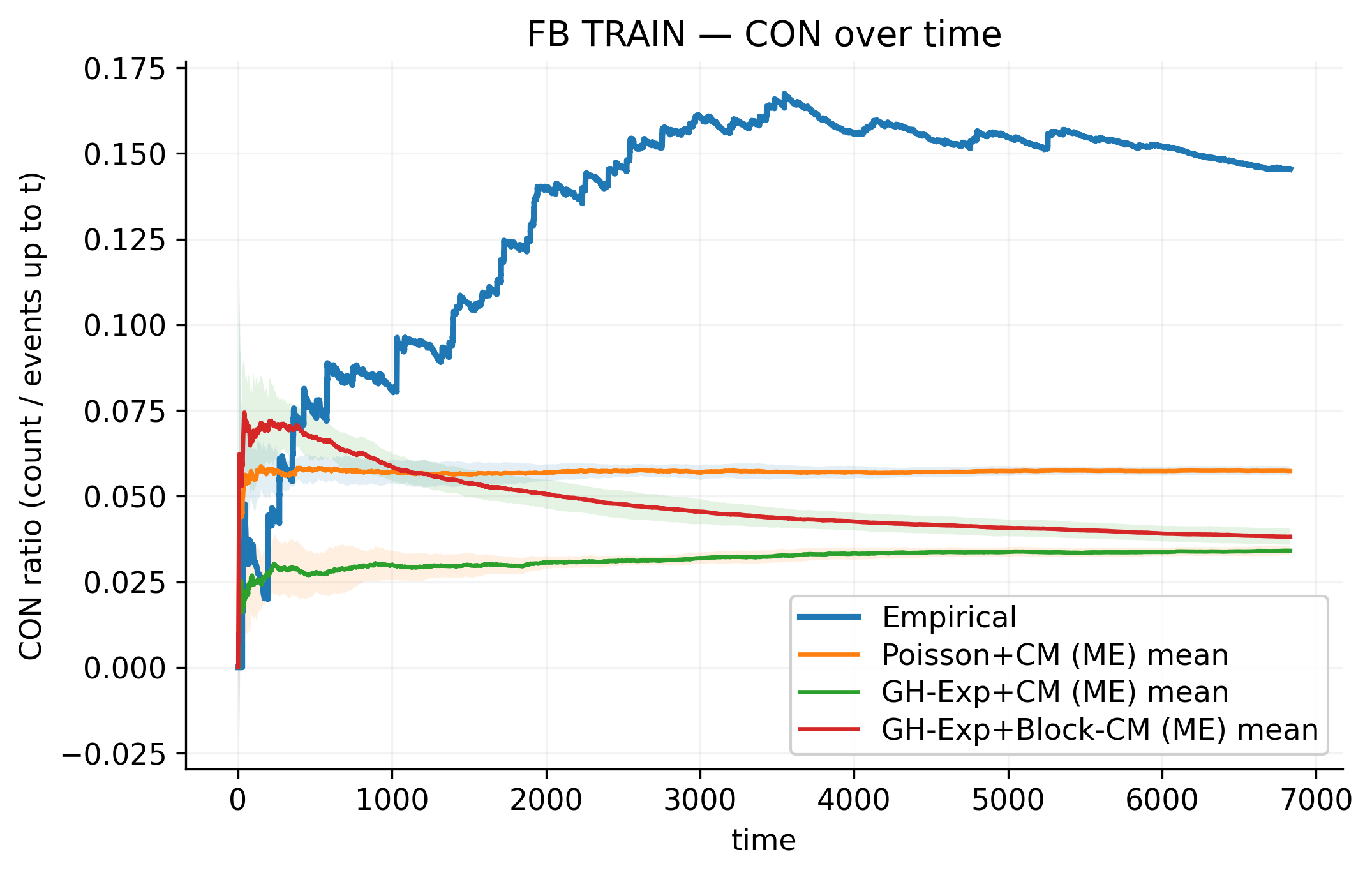}
    \caption{CON, $\Delta_{\text{large}}$}
  \end{subfigure}\hfill
  \begin{subfigure}{0.48\textwidth}
    \includegraphics[width=\linewidth]{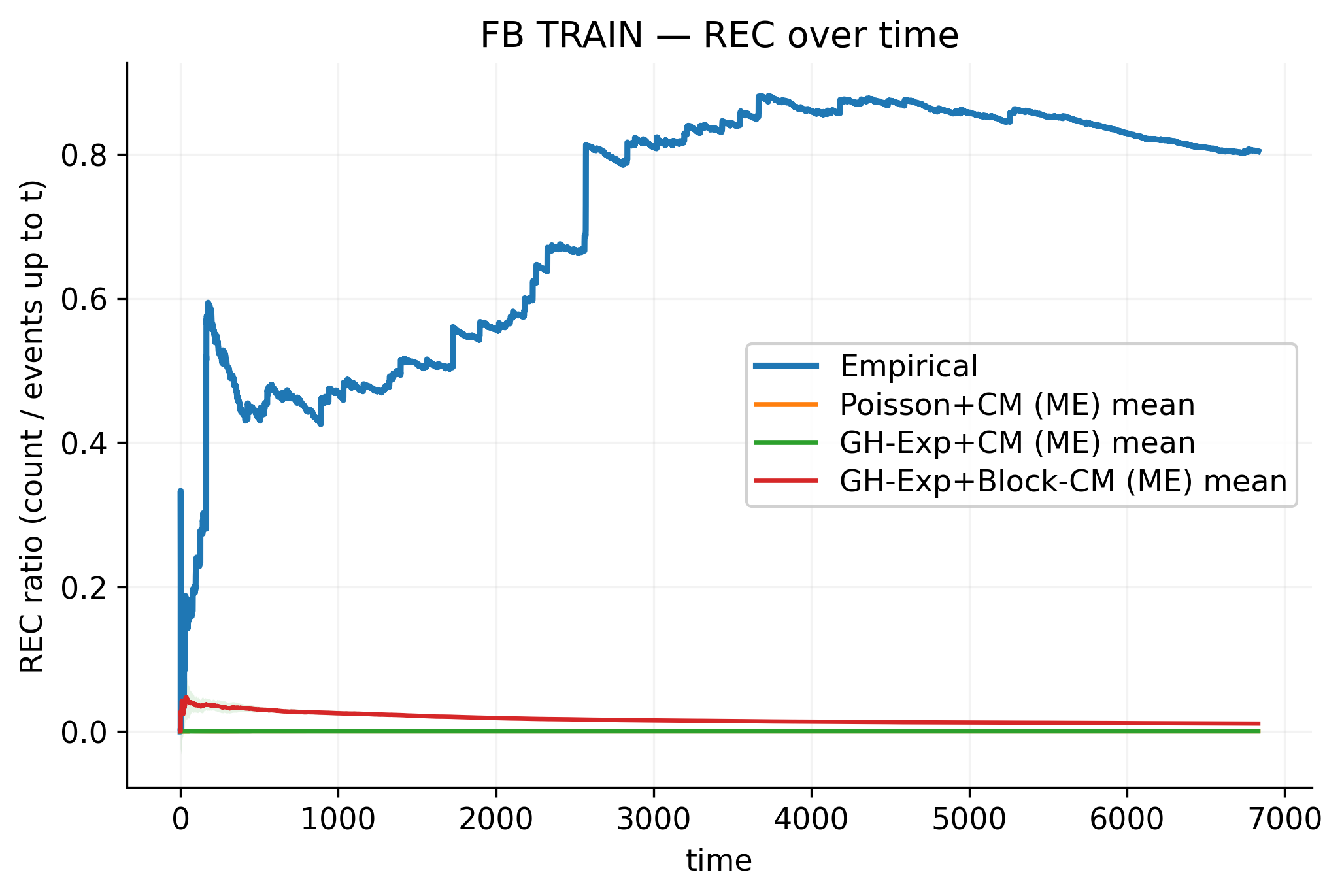}
    \caption{REC, $\Delta_{\text{large}}$}
  \end{subfigure}
  \vspace{0.75em}
  \begin{subfigure}{0.48\textwidth}
    \includegraphics[width=\linewidth]{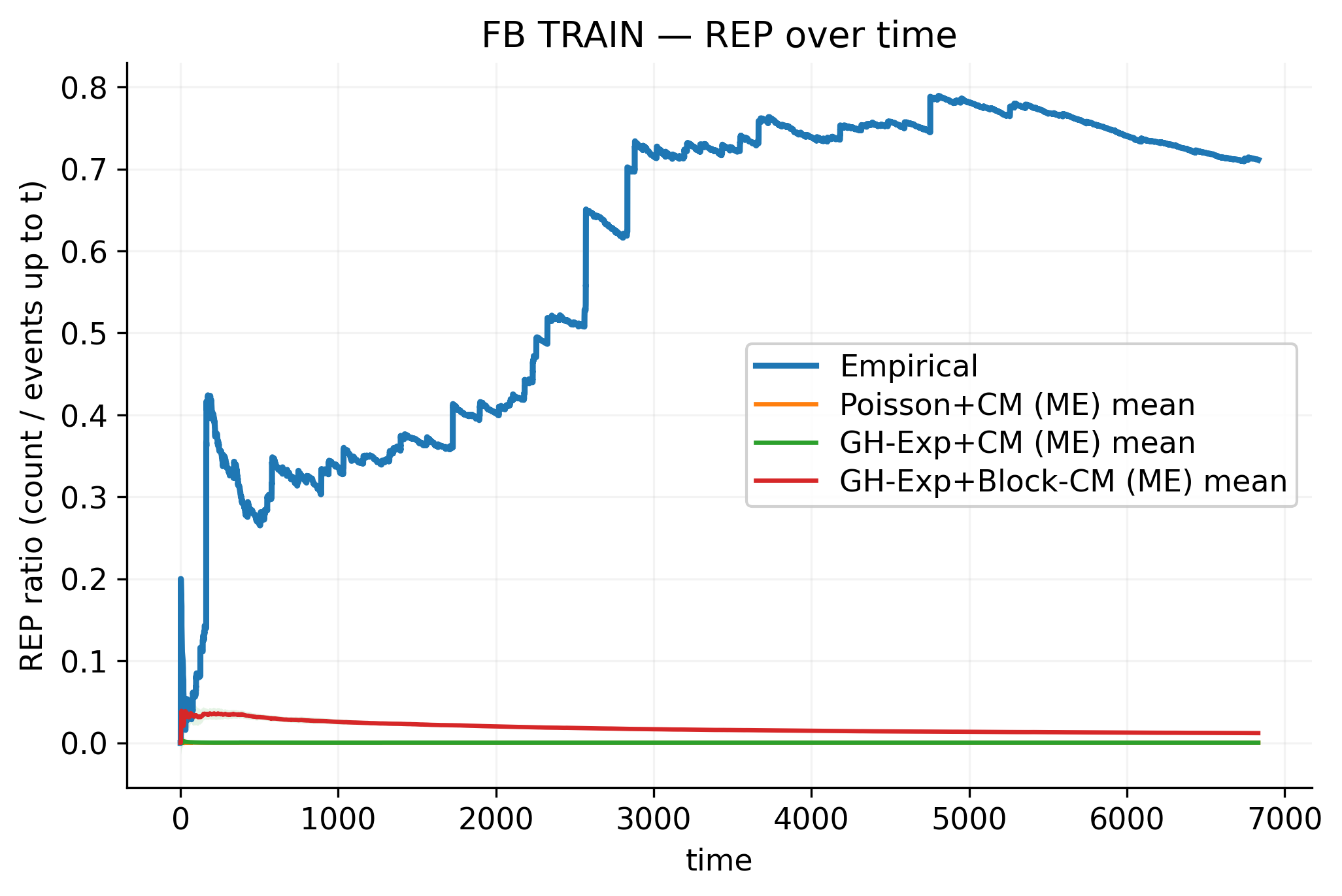}
    \caption{REP, $\Delta_{\text{large}}$}
  \end{subfigure}\hfill
  \begin{subfigure}{0.48\textwidth}
    \includegraphics[width=\linewidth]{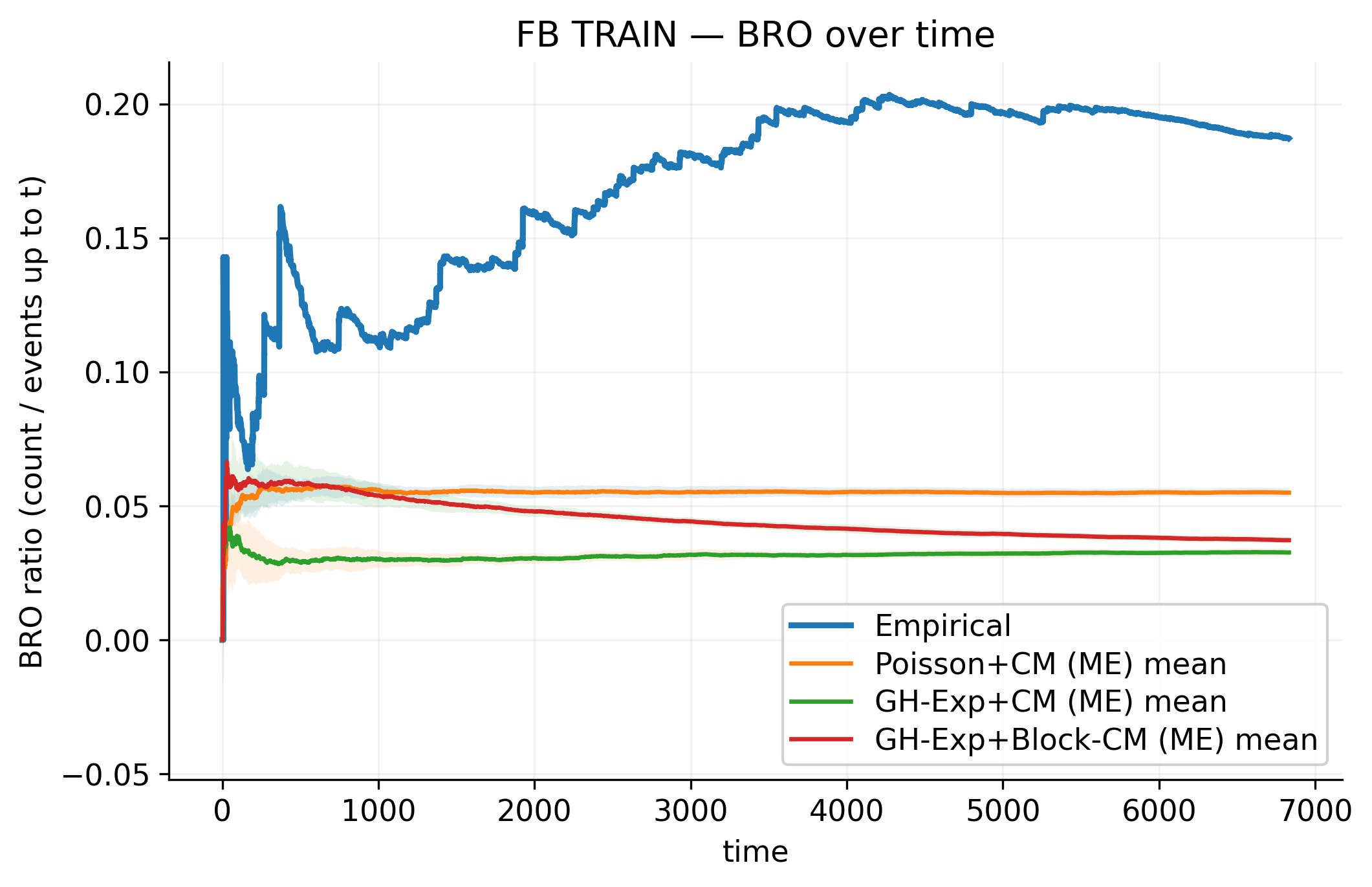}
    \caption{BRO, $\Delta_{\text{large}}$}
  \end{subfigure}
  \caption{Facebook TRAIN motif ratios (mean $\pm$ s.d.) over time for $\Delta_{\text{large}}$.}
  \label{fig:fb-motifs-large}
\end{figure*}

We here provide the motif analysis for the four benchmark temporal networks. 
These datasets span domains from organizational communication to international relations, physical proximity, and online social media. 
They also differ in density, timescale, and heterogeneity.
The Enron email corpus consists of time-stamped internal communications among employees of the Enron Corporation over multiple years. It is widely used as a benchmark dataset for temporal network modeling due to its marked structural heterogeneity (e.g., hubs and community organization) and pronounced burstiness in communication activity , see Fig.\ref{fig:enron-motifs-large}.
The Militarized Interstate Disputes (MID) database captures time-stamped interactions among states in the international system, recording conflicts and confrontations, see Fig. \ref{fig:mid-motifs-large}. 
This network represents a very different context, where interactions are sparse, heterogeneous in scale, and subject to geopolitical constraints. 
The third is the Reality Mining dataset, which measures proximity interactions among participants in an MIT community through mobile devices. Unlike the email and conflict data, Reality Mining captures face-to-face encounters at high temporal resolution, with strong daily and weekly rhythms superimposed on bursty group interactions , see Fig.\ref{fig:reality-motifs-large}.
Finally, the Facebook dataset provides a record of social media activity, in which messages between users form a large-scale online temporal communication network characterized by heavy-tailed degree distributions and extreme temporal clustering, see Fig.\ref{fig:fb-motifs-large}.
In the Enron dataset, we find that reciprocated motifs occur significantly more frequently than predicted by the maximum-entropy baseline conditioned on node strengths and Hawkes-driven bursty timing.
In the MID dataset, one can assess whether conflict proliferation across specific edges or blocks is more common than expected given the observed distribution of disputes. 
In Reality Mining, deviations from the null highlight the role of small-group gatherings and scheduled rhythms, while in Facebook the null benchmark reveals when extreme broadcast behavior of hubs is attributable to degree heterogeneity alone or signals additional bursty mechanisms at a node level.

\begin{figure*}[t]
  \centering
  \begin{subfigure}{0.48\textwidth}
    \includegraphics[width=\linewidth]{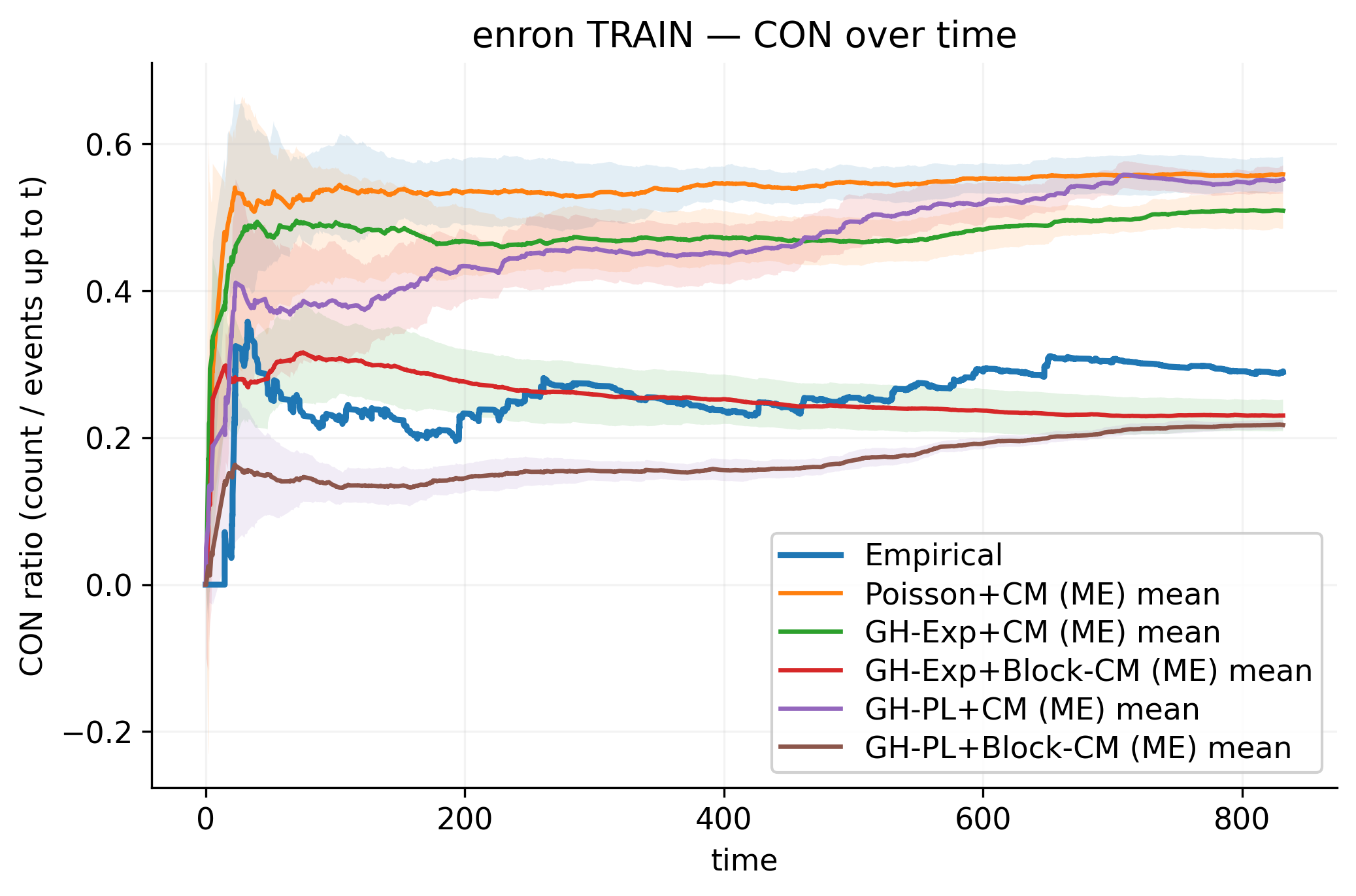}
    \caption{CON, $\Delta_{\text{large}}$}
  \end{subfigure}\hfill
  \begin{subfigure}{0.48\textwidth}
    \includegraphics[width=\linewidth]{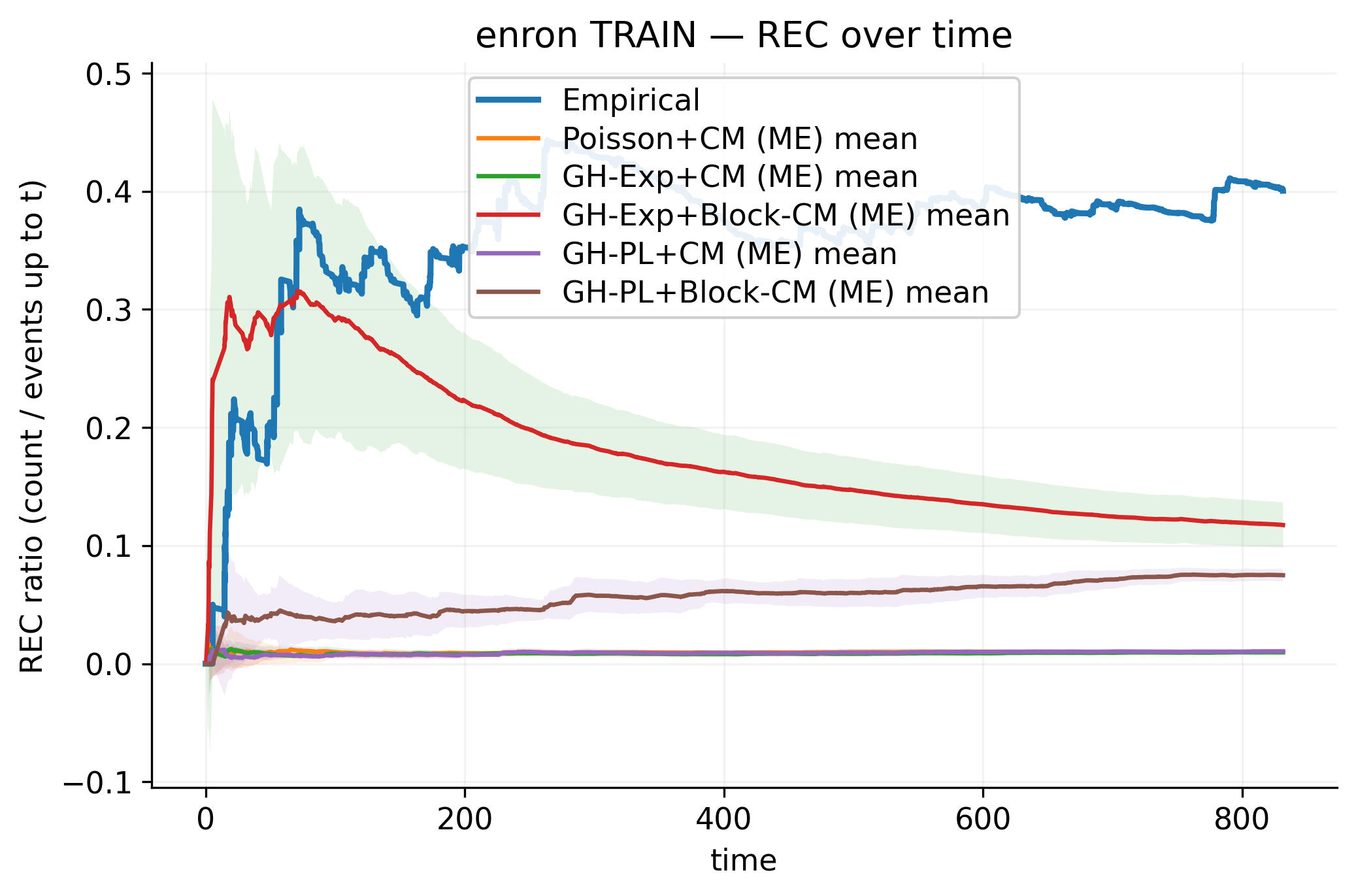}
    \caption{REC, $\Delta_{\text{large}}$}
  \end{subfigure}
  \vspace{0.75em}
  \begin{subfigure}{0.48\textwidth}
    \includegraphics[width=\linewidth]{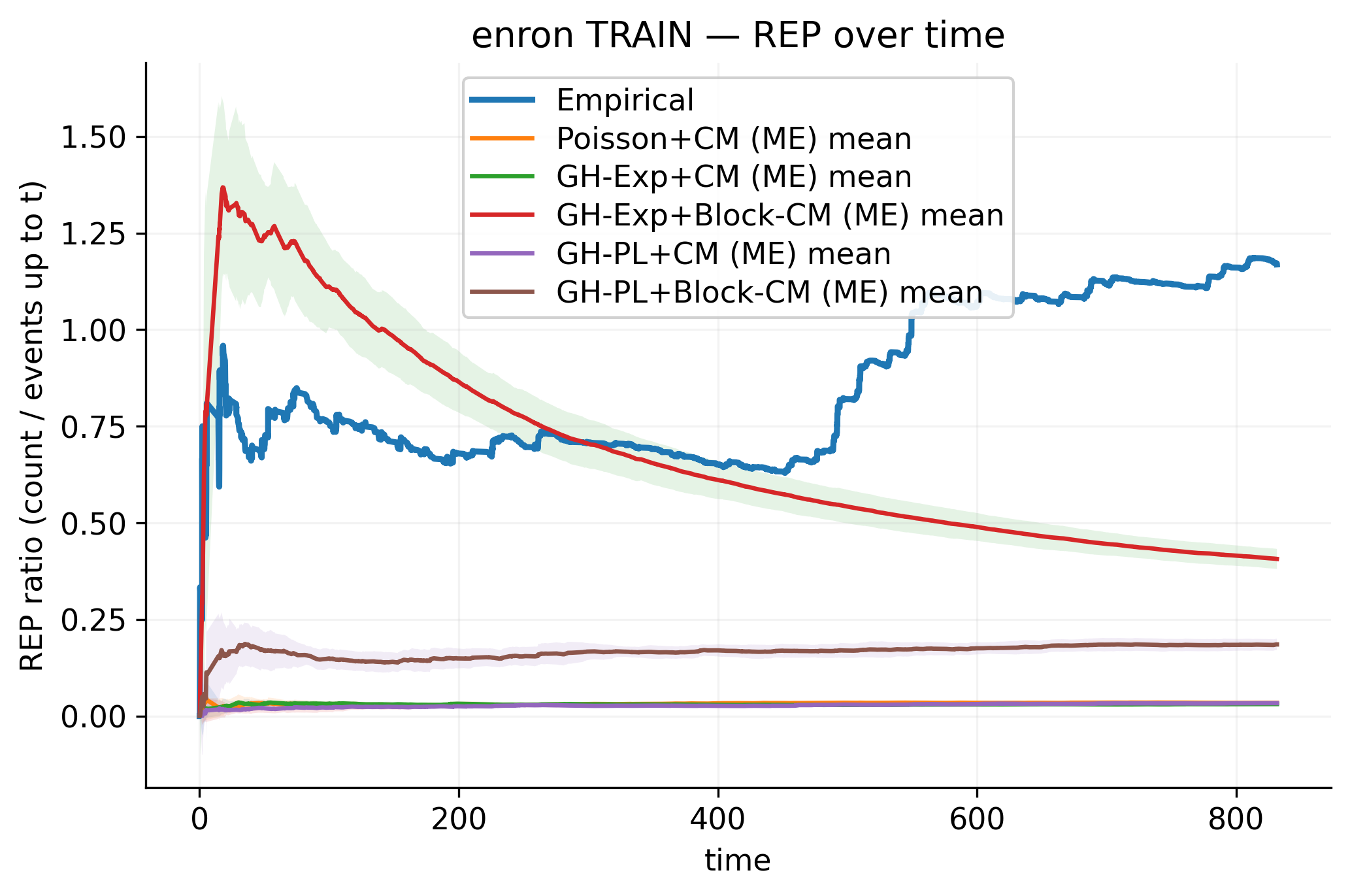}
    \caption{REP, $\Delta_{\text{large}}$}
  \end{subfigure}\hfill
  \begin{subfigure}{0.48\textwidth}
    \includegraphics[width=\linewidth]{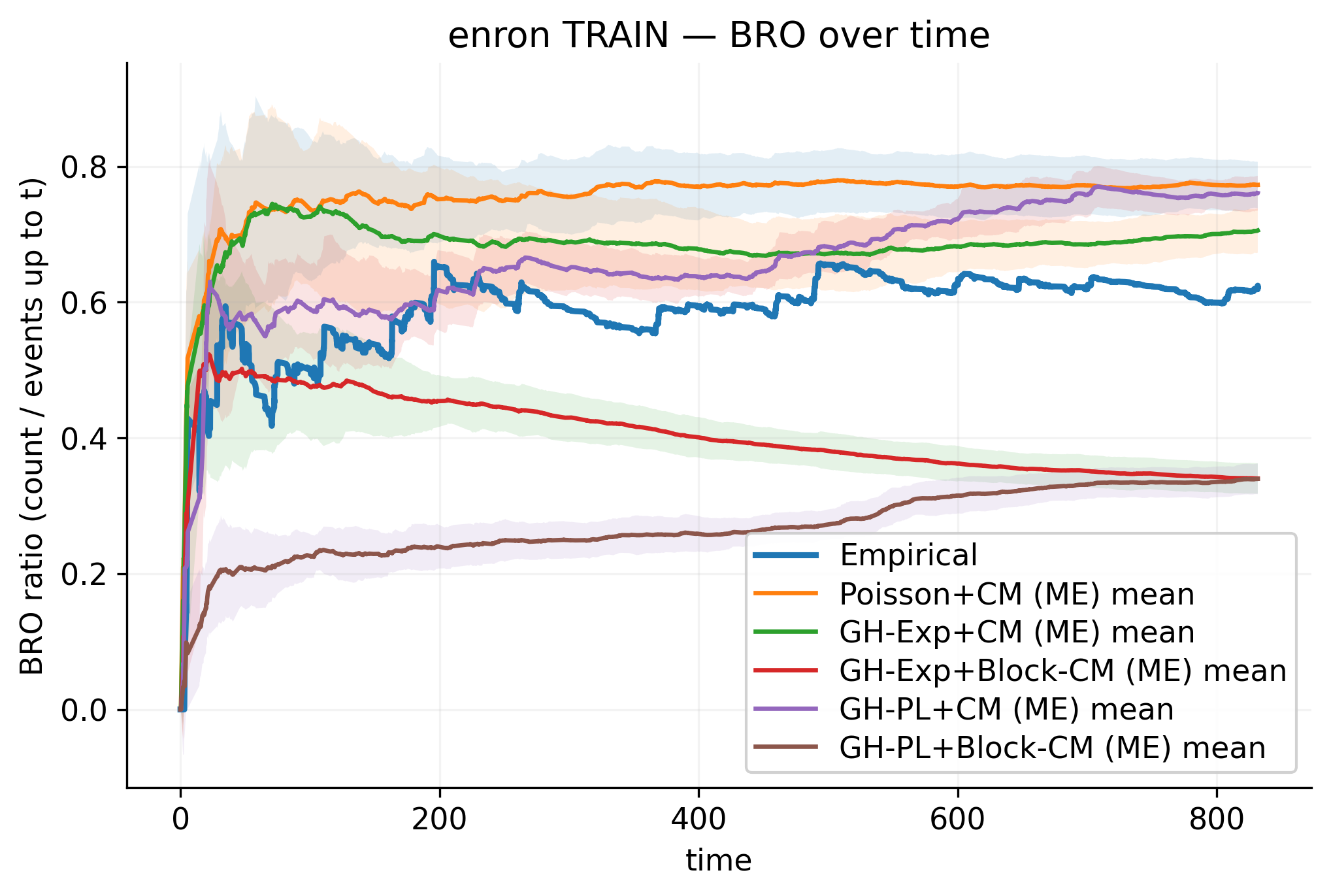}
    \caption{BRO, $\Delta_{\text{large}}$}
  \end{subfigure}
  \caption{Enron TRAIN motif ratios (mean $\pm$ s.d.) over time for $\Delta_{\text{large}}$.}
  \label{fig:enron-motifs-large}
\end{figure*}

\begin{table}[!h]
\centering
\caption{Per-event log-likelihoods for the \textbf{Enron} dataset under different temporal models. 
LL\textsubscript{mark/ev} refers to the mark (dyad) likelihood, 
LL\textsubscript{time/ev} to the temporal likelihood, and LL\textsubscript{total/ev} to their sum.}
\label{tab:ll-enron}
\begin{tabular}{llrrr}
\toprule
\textbf{Partition} & \textbf{Time model} 
& \textbf{LL\textsubscript{mark/ev}} 
& \textbf{LL\textsubscript{time/ev}} 
& \textbf{LL\textsubscript{total/ev}} \\
\midrule
Global    & Poisson     & $-7.53$ & $0.283$ & $-7.25$ \\
Global    & Hawkes-Exp  & $-7.53$ & $0.291$ & $-7.24$ \\
Global    & Hawkes-PL   & $-7.53$ & $0.775$ & $-6.76$ \\
Blockpair & Poisson     & $-3.45$ & $-2.10$ & $-5.55$ \\
Blockpair & Hawkes-Exp  & $-3.45$ & $-2.06$ & $-5.50$ \\
Blockpair & Hawkes-PL   & $-3.45$ & $-1.61$ & $-5.06$ \\
\bottomrule
\end{tabular}
\end{table}


\begin{table*}[t]
\centering
\resizebox{\linewidth}{!}{
\begin{tabular}{lrrrrrr}
\toprule
\textbf{Model} & \textbf{Nodes} & \textbf{Events} & \textbf{Unique Edges} & \textbf{ISI Mean} & \textbf{CV} & \textbf{Burstiness} \\
\midrule
Empirical & 136 & 2999 & 513 & 0.277 & 2.74 & 0.465 \\
\midrule
Poisson+CM & $135.5 \pm 0.5$ & $2990 \pm 60$ & $1450 \pm 20$ & $0.279 \pm 0.005$ & $0.991 \pm 0.016$ & $-0.004 \pm 0.008$ \\
GH-Exp+CM & $135.3 \pm 0.5$ & $2700 \pm 60$ & $1350 \pm 30$ & $0.308 \pm 0.007$ & $1.00 \pm 0.02$ & $0.001 \pm 0.011$ \\
GH-Exp+Block-CM & $135.9 \pm 0.3$ & $2600 \pm 50$ & $697 \pm 16$ & $0.320 \pm 0.006$ & $1.022 \pm 0.018$ & $0.011 \pm 0.009$ \\
GH-PL+CM & $135.4 \pm 0.5$ & $2700 \pm 40$ & $1349 \pm 18$ & $0.309 \pm 0.005$ & $1.23 \pm 0.02$ & $0.105 \pm 0.009$ \\
GH-PL+Block-CM & $135.6 \pm 0.5$ & $2510 \pm 50$ & $697 \pm 11$ & $0.331 \pm 0.007$ & $1.04 \pm 0.03$ & $0.020 \pm 0.013$ \\
\bottomrule
\end{tabular}
}
\caption{Enron TRAIN: model summary. 
For each model, the table reports the number of nodes, total number of events, number of unique directed edges, mean inter-event time (ISI\_mean), coefficient of variation (CV), and burstiness index. 
Values are shown as mean ± standard deviation across generated samples.}
\label{tab:enron-model-summary}
\end{table*}


\begin{table*}[t]
\centering
\resizebox{0.8\linewidth}{!}{
\begin{tabular}{lrrrr}
\toprule
\textbf{Model} & \textbf{REP} & \textbf{REC} & \textbf{CON} & \textbf{BRO} \\
\midrule
Empirical & 1.17 & 0.40 & 0.29 & 0.624 \\
\midrule
Poisson+CM & $0.034 \pm 0.004$ & $0.010 \pm 0.002$ & $0.56 \pm 0.02$ & $0.77 \pm 0.04$ \\
GH-Exp+CM & $0.031 \pm 0.004$ & $0.0096 \pm 0.0020$ & $0.51 \pm 0.03$ & $0.71 \pm 0.03$ \\
GH-Exp+Block-CM & $0.41 \pm 0.03$ & $0.12 \pm 0.02$ & $0.23 \pm 0.02$ & $0.34 \pm 0.02$ \\
GH-PL+CM & $0.033 \pm 0.003$ & $0.0105 \pm 0.0019$ & $0.552 \pm 0.020$ & $0.76 \pm 0.03$ \\
GH-PL+Block-CM & $0.185 \pm 0.015$ & $0.075 \pm 0.006$ & $0.217 \pm 0.007$ & $0.34 \pm 0.02$ \\
\bottomrule
\end{tabular}
}
\caption{Temporal motif ratios (Enron TRAIN, $\Delta_{\text{large}}$). 
Ratios are computed as the normalized counts of reciprocal (REP), repeated (REC), convergent (CON), and broadcast (BRO) motifs within the large temporal window $\Delta_{\text{large}}$.}
\label{tab:enron-motif-summary}
\end{table*}

\end{document}